\documentclass[11pt]{article}
\usepackage{comment}
\usepackage[margin=2.5cm]{geometry}
\usepackage{amsmath,amssymb,extarrows,mathtools,graphicx,subfigure,setspace,bbold}
\usepackage{cite}
\usepackage{slashed}
\usepackage[dvipsnames]{xcolor}
\usepackage{hyperref}
\hypersetup{colorlinks=true, linkcolor=blue, citecolor=magenta, linktoc=page}
\usepackage{tikz}
\usepackage{arydshln,leftidx,mathtools}
\usetikzlibrary{decorations.pathreplacing}
\numberwithin{equation}{section}
\newcommand{\be}{\begin{equation}}
\newcommand{\bea}{\begin{eqnarray}}
\newcommand{\eea}{\end{eqnarray}}
\newcommand{\ba}{\begin{align}}
\newcommand{\ea}{\end{align}}
\newcommand{\ee}{\end{equation}}

\begin{document}

\begin{titlepage}
\thispagestyle{empty}

\begin{flushright}
IPM/P-2016/032\\
\end{flushright}

\vspace{.4cm}
\begin{center}
\noindent{\Large \textbf{Non-local Probes in\\ \vspace{2mm} Holographic Theories with Momentum Relaxation}}\\
\vspace{2cm}

M. Reza Mohammadi Mozaffar$^a$, Ali Mollabashi${}^{a}$, Farzad Omidi${}^{b}$
\vspace{1cm}

${}^a$ {\it School of Physics,} ${}^b$ {\it School of Astronomy,}
\\
{\it Institute for Research in Fundamental Sciences (IPM), Tehran, Iran}\\
\vspace{1cm}
Emails: {\tt m$_{-}$mohammadi, mollabashi, and farzad@ipm.ir}

\vskip 2em
\end{center}

\vspace{.5cm}
\begin{abstract}
We consider recently introduced solutions of Einstein gravity with minimally coupled massless scalars. The geometry is homogeneous, isotropic and asymptotically anti de-Sitter while the scalar fields have linear spatial-dependent profiles. The spatially-dependent marginal operators dual to scalar fields cause momentum dissipation in the deformed dual CFT. We study the effect of these marginal deformations on holographic entanglement measures and Wilson loop. We show that the structure of the universal terms of entanglement entropy for $d>2$-dim deformed CFTs is corrected depending on the geometry of the entangling regions. In $d=2$ case, the universal term is not corrected while momentum relaxation leads to a non-critical correction. We also show that decrease of the correlation length causes: the phase transition of holographic mutual information to happen at smaller separations and the confinement/deconfinement phase transition to take place at smaller critical lengths. The effective potential between point like external objects also gets corrected. We show that the strength of the corresponding force between these objects is an increasing function of the momentum relaxation parameter.
\end{abstract}

\end{titlepage}

\newpage

\tableofcontents
\noindent
\hrulefill

\onehalfspacing

\section{Introduction}
A great amount of interests and attempts have been dedicated to understand strongly interacting systems in the context of AdS/CMT (see \cite{Hartnoll:2011fn ,Iqbal:2011ae} for reviews). Generically different states of such systems are described in terms of solutions of Einstein-Maxwell-Dilaton (EMD) theories. Although these theories seem to reproduce a family of essential features of such systems, there exists very important features which are not captured by solutions of EMD theories as gravity duals of such theories.

Solutions of EMD theories have a net amount of charge and are fully translational invariant. In such a case applying a tiny electric field is enough to result in an infinite DC conductivity. This is not what is known from realistic systems, thus the gravity dual needs some improvements. To overcome such a feature and find the expected Drude behavior, people have proposed several ways to provide mechanisms for the charge carriers to relax their momentum. To our knowledge, this is done either by considering probe objects \cite{Probe} or breaking the translational symmetry of the system. Breaking the translational symmetry itself can be done within different mechanisms. This is studied either by considering impurities in the system \cite{bts1}, breaking the diffeomorphism invariance in the bulk theory \cite{bts2}, turning on spatial dependent sources \cite{bts3, Andrade:2013gsa} or considering backreacted geometries from probe charged matter \cite{bts4}.

Here we are interested in a specific family of models which have spatially dependent sources. The model of our interest is what was first introduced by Andrade and Withers in \cite{Andrade:2013gsa}. The transport properties and various generalizations (in different directions) of this family of models have been widely studied in the literature. The idea is a very simple one: in order to break momentum conservation, one may consider a (number of) spatial-dependent scalar field(s) in the bulk thus the Ward identity takes the following form
\be
\nabla^iT_{ij}+\langle\mathcal{O}_\chi\rangle\partial_{j} \chi=0.
\ee
Note that these models are usually considered in presence of a gauge field in the bulk theory which has a position dependent source $A_i$. In this case there would also be a term due to the gauge field as $\langle J^i\rangle F_{ij}$ in the Ward identity. Here since we are not interested in any transport coefficient, we will not consider any gauge field in our model.

Andrade-Withers model which we review in the next section is composed of gravity and some minimally coupled massless scalar fields. The essential point is that since this theory admits solutions with linear spatial-dependence of the scalar field profiles, the contribution of the scalars to the stress tensor is homogeneous and together with considering $(d-1)$ scalars (the number of spatial dimensions of the dual field theory) one can engineer homogeneous and isotropic black-brane solutions.

As we have mentioned earlier, we consider black-brane solutions which are neutral.
Moreover, we would like to emphasis that we are mainly interested in considering massless black-branes where the event horizon is caused merely by the momentum dissipation parameter.
Such solutions could be found either in Andrade-Withers model, which are sometimes called \textit{polynomial models} or even in a more strange family of models introduced by Taylor and Woodhead in \cite{Taylor:2014tka} where the scalar fields are under square root in the action. We will mainly consider polynomial models in this paper and report some features of square root models in the discussion section.

In these models which we consider, the massless scalar fields are dual to marginal operators in the dual field theory. These marginal operators do not affect the UV structure of the dual theory but have non-trivial subleading effects in the holographic RG flow. The goal of this paper is to study the momentum dissipation effects on holographic non-local measures such as entanglement entropy and Wilson loop.
The geometries which we are interested in, having non-vanishing momentum dissipation parameter, are interpreted as new vacuum states in the dual theory which we call ``non-conformal vacuums"
Having this in mind, in this paper we often consider the momentum relaxation parameter perturbatively just for simplicity of our analysis. 
Furthermore to avoid mixture of thermal and quantum effects, in some parts we also consider solutions with non-vanishing mass, to study holographic entanglement entropy in extremal geometries which is dual to a zero temperature (but of course mixed) states.

The outlook of this paper is as follows: in section \ref{sec:setup} we introduce the model of our interest and some essential properties of it. Sections \ref{sec:HEE} and \ref{sec:other} are dedicated to holographic study of entanglement measures including entanglement entropy and mutual information. We continue in section \ref{sec:geo} by investigating the momentum relaxation effects on the phase transition of geometric entropy. Moreover, in section \ref{sec:wilson} we study the effective potential between point like external objects in such theories using holographic Wilson loop. In the last section we make our concluding remarks.

\section{Holographic Theories with Momentum Relaxation}\label{sec:setup}

In this section we introduce specific holographic models of our interest which are dual to quantum field theories in presence of momentum relaxation. As we have mentioned in the introduction section, there are several families of such models. Here in the general family of models with spatial dependent sources, we mainly consider one specific simple one.

This model which we sometimes refer to it by the polynomial model is defined in $(d+1)$-dimensions by the following action \cite{Andrade:2013gsa}\footnote{Since one important feature of systems with momentum relaxation is the so-called Drude behavior of the DC conductivity, the authors of \cite{Andrade:2013gsa} have considered a gauge field in this model in order to verify such a behavior. Here since we are not interested in studying any transport coefficient of this model, we turn off the gauge field from the very beginning of our analysis.}
\bea\label{polyaction}
I=\frac{1}{16 \pi G_N}\int d^{d+1}x \sqrt{-g}\left[\mathcal{R}-2\Lambda
-\frac{1}{2}\sum_{I=1}^{d-1}(\partial\chi_I)^2\right],
\eea
where $\Lambda=-\frac{d(d-1)}{2L^2}$, 
and $\chi_I$'s are massless scalar fields. Here $I$ is an internal index denoting the $(d-1)$ scalar fields. This action has an asymptotically AdS$_{d+1}$ black-brane solution with a non-trivial profile for the scalar fields (for $d>2$) as follows
\begin{align}\label{staticmetric}
\begin{split}
ds^2&=\frac{L^2}{\rho^2}\left[-f(\rho)dt^2+\frac{d\rho^2}{f(\rho)}+dx_{d-1}^2\right],\\
f(\rho)&=1-\frac{\alpha^2 \rho^2}{2(d-2)}-m_0 \rho^d,\\
\chi_I(x^a)&=\alpha_{Ia} x^a,
\end{split}
\end{align}
where $a$ denotes the $d-1$ spatial directions and 
\be
\alpha^2\equiv \frac{1}{d-1}\sum_{a=1}^{d-1}\vec{\alpha}_a.\vec{\alpha}_a, \;\;\;(\vec{\alpha}_a)_I=\alpha_{Ia}.
\ee
The scalar fields are dimensionless and $\alpha_{Ia}$'s have dimension of inverse length.
Note that for $d=2$ the solution reads as
\begin{align}\label{2dmetric}
\begin{split}
f(\rho)&=1+\frac{\alpha^2 \rho^2}{2}\log \rho-m_0 \rho^2,\\
\chi(x)&=\alpha x.
\end{split}
\end{align}
The temperature of the black-brane is given by
\bea
T=\frac{d}{4\pi\rho_h}\left(1-\frac{\alpha^2\rho_h^2}{2d}\right),\;\;\;m_0=\frac{1}{\rho_h^d}\left(1-\frac{\alpha^2\rho_h^2}{2(d-2)}\right),
\eea
which for $d>2$  with
\be\label{eq:PolyExt}
\alpha^2=\frac{2d}{\rho_h^2},\;\;\;\;\;\;\;\;m_0=\frac{2}{2-d}\frac{1}{\rho_h^d},
\ee
and for $d=2$ with the following choice of the parameters
\be\label{eq:PolyExt2}
\alpha^2=\frac{4}{\rho_h^2},\;\;\;\;\;\;\;\;m_0=\frac{1+2\log\rho_h}{\rho_h^2},
\ee
leads to an extremal black-brane where $f(\rho_h)=f'(\rho_h)=0$. Here the extremal solution exists due to the momentum relaxation parameter rather than a $U(1)$ charge in comparison with the case of RN-AdS black-brane. Also note that in the near horizon limit, considering the following scaling limit for $\lambda\rightarrow 0$
\bea
\rho-\rho_h=\frac{\rho_h^2}{d\xi}\lambda,\;\;\;t=\frac{\tau}{\lambda},
\eea
the resultant near horizon region is an AdS$_2\times \mathbb{R}^{d-1}$, which is
\bea
ds^2=\frac{L_2^2}{\xi^2}\left(-d\tau^2+d\xi^2\right)+\frac{\alpha^2L_2^2}{2}dx_{d-1}^2,\;\;\;L_2^2=\frac{L^2}{d}. 
\eea

\subsection*{Hyperscaling Violating Generalization}
An interesting generalization of the polynomial model is to consider asymptotically non-relativistic backgrounds which have non-trivial dynamical and hyperscaling violating exponents, $z$ and $\theta$. These kind of solutions are constructed by adding
some axion fields to the EMD theories, and has been studied recently in \cite{Ge:2016lyn}\footnote{For other types of anisotropic hyperscaling violating solutions see \cite{Roychowdhury:2015fxf, Cremonini:2016avj}.} with the following action\footnote{Note that in this paper whenever we discuss about hyperscaling violating solutions we consider $(d+2)$-dimensional gravity solutions thus $(d+1)$-dimensional dual field theories.}
\be
I=\frac{1}{16 \pi G_N}\int d^{d+2}x \sqrt{-g}\left[\mathcal{R}+V(\phi)-\frac{1}{2}(\partial \phi)^2-\frac{1}{4}Z(\phi)F_{\mu\nu}F^{\mu\nu}
-\frac{1}{2}Y(\phi)\sum_{I=1}^{d}(\partial\chi_I)^2\right],
\ee
where $Z(\phi)=e^{\lambda_1\phi}$ and $Y(\phi)=e^{-\lambda_2\phi}$. The corresponding solution is given by
\begin{alignat}{2}\label{hsmetric}
\begin{split}
ds^2&=\rho^{\frac{2(\theta-d)}{d}}\left[-\frac{f(\rho)}{\rho^{2(z-1)}}dt^2+\frac{d\rho^2}{f(\rho)}+d\vec{x}^2_d\right],\;\;\;\;\;
f(\rho)=1-m_0\rho^{d+z-\theta}-\alpha^2\rho^{2(z-\frac{\theta}{d})},
\end{split}
\end{alignat}
together with
\begin{align}
\begin{split}
F_{\rho t}&=\sqrt{2(z-1)(z+d-\theta)}\rho^{1+\theta-d-z}, \;\;\;\;\;\;
\phi =-\sqrt{2(d-\theta)(z-1-\theta/d)}\ln \rho,\\
\chi_I(x^a)&=\alpha_{Ia} x^a,\hspace{41mm}
V(\phi)=(z+d-\theta-1)(z+d-\theta)\rho^{\frac{-2\theta}{d}},
\end{split}
\end{align}
where
\begin{align}
\begin{split}
\lambda_1 &=\frac{\sqrt{2}(\theta-d-\theta/d)}{\sqrt{(d-\theta)(z-1-\theta/d)}},\;\;\;\;\;\;\;\;\;\;\;\;\;\;\;
\lambda_2 =-\sqrt{2\frac{z-1-\theta/d}{d-\theta}}\\
\alpha^2 &=\frac{d^2\alpha_0^2}{2(d-\theta)(d^2+2\theta-d(z+\theta))},\;\;\;\;\;\;
\alpha_0^2 \equiv \frac{1}{d}\sum_{a=1}^{d}\vec{\alpha}_a.\vec{\alpha}_a.
\end{split}
\end{align}

\section{Holographic Entanglement Entropy}\label{sec:HEE}

A natural question about such marginal deformations in the field theory would be how entanglement entropy is affected due to these types of deformations? Entanglement entropy is believed to capture some universal information about the field theory such as anomaly coefficients of the stress tensor and also some information about their behavior under renormalization group flow at least in certain cases. Since we are interested in deformed states of CFTs which are dual to asymptotically AdS geometries, here in this section we are going to use Ryu-Takayanagi holographic proposal \cite{Ryu:2006ef,Ryu:2006bv} to study entanglement entropy as a probe of how momentum relaxation caused due to specific marginal deformations may affect the UV CFT.\footnote{Here we would like to note that to our knowledge there are two related studies in the literature which are \cite{Azeyanagi:2009pr} and \cite{Banerjee:2014oaa}. The authors of these papers have briefly studied holographic entanglement entropy in \textit{anisotropic} models with momentum relaxation.}

In what follows in this section we study holographic entanglement entropy (HEE) in the model introduced in \eqref{staticmetric}. This is done for different entangling regions to investigate the role of momentum relaxation (marginal deformation of the CFT) on the HEE. We consider infinite strip, spherical and cylindrical entangling regions defined as below.

For strip entangling region we have $dx_{d-1}^2=\sum_{i=1}^{d-1}dx_i^2$. The entangling region is defined as
\be
-\frac{\ell}{2}\leq x_1\equiv x \leq \frac{\ell}{2},\;\;\; -\frac{H}{2}\leq x_{i>1}\leq \frac{H}{2},\;\;\;H\gg \ell.
\ee

For spherical entangling region we have $dx_{d-1}^2=dr^2+r^2 d\Omega_{d-2}^2$. The entangling region is defined as $0<r<\ell$.

For cylindrical entangling region we have $dx_{d-1}^2=du^2+dr^2+r^2 d\Omega_{d-3}^2$ where $u$ is the coordinate along the height direction of the cylinder. The entangling region is defined as 
\be
0<r<\ell,\;\;\; 0<u<H,\;\;\;H\gg \ell.
\ee
Also in the following sections we will study some other entanglement measures including holographic mutual information, information metric and phase transitions of double wick-rotated solutions. 

\subsection{Strip Entangling Region}

Considering the geometry \eqref{staticmetric}, the corresponding hypersurface can be parametrized as $x=x(\rho)$ and the induced metric on the hypersurface is given by
\be
ds^2_{\rm ind.}=\frac{L^2}{\rho^2}\left[\left({x'}^2+\frac{1}{f(\rho)}\right)d\rho^2+\sum_{i=2}^{d-1}dx_i^2\right],
\ee
where prime denotes the derivative with respect to $\rho$. Using the above expression the area of the corresponding hypersurface is given by
\bea
\mathcal{A}=L^{d-1}H^{d-2}\int \frac{d\rho}{\rho^{d-1}}\sqrt{{x'}^2+\frac{1}{f(\rho)}}.
\eea
This functional dose not depend on $x(\rho)$ explicitly and the equation of motion leads to
\bea
x'(\rho)=\frac{1}{\sqrt{\left(\frac{\rho_t^{2(d-1)}}{\rho^{2(d-1)}}-1\right)f(\rho)}},
\eea
where $\rho_t$ is the turning point of the hypersurface with $x'(\rho_t)=\infty$. In this case the length of the strip and the area of the minimal hypersurface are given by
\begin{align}\label{lA1}
\begin{split}
\ell&=2\int_0^{\rho_t} d\rho\frac{1}{\sqrt{\left(\frac{\rho_t^{2(d-1)}}{\rho^{2(d-1)}}-1\right)f(\rho)}},\\
\mathcal{A}&=2L^{d-1}H^{d-2}\int_0^{\rho_t} \frac{d\rho}{\rho^{d-1}}\frac{1}{\sqrt{\left(1-\frac{\rho^{2(d-1)}}{\rho_t^{2(d-1)}}\right)f(\rho)}}.
\end{split}
\end{align}
The above integrals do not have analytic results in arbitrary dimension, therefore we consider different specific cases as follows:

\subsubsection*{(i) Case $m_0=0$ and $\alpha\ell\ll 1$}
Here since we are considering  $\alpha\ell$ as a small parameter, and for fixed $\ell$ in the $\alpha\to0$ limit we are left with the results in the pure AdS case, we report the $\alpha$-dependent part of the HEE as 
\begin{align}\label{stripHEE}
\begin{split}
\Delta S&=\frac{L^{d-1}H^{d-2}}{8G_N(d-2)}\left[\frac{1}{(d-4)\epsilon^{d-4}}+\frac{2^{d-5}\sqrt{\pi}}{3\ell^{d-4}}\left(\frac{\sqrt{\pi}\Gamma\left(\frac{d}{2(d-1)}\right)}{\Gamma\left(\frac{1}{2(d-1)}\right)}\right)^{d-4}\frac{\Gamma\left(\frac{4-d}{2(d-1)}\right)}{\Gamma\left(\frac{3}{2(d-1)}\right)}\right]\alpha^2+\mathcal{O}\left(\alpha^4\right),
\end{split}
\end{align}
where $\Delta S=S-S_0$ and $S_0$ is the HEE for the $\alpha=0$ case which is given by
\begin{align}
\begin{split}
S_0&=\frac{L^{d-1}H^{d-2}}{2G_N(d-2)}\left[\frac{1}{\epsilon^{d-2}}-\frac{2^{d-2}}{\ell^{d-2}}\left(\frac{\sqrt{\pi}\Gamma\left(\frac{d}{2(d-1)}\right)}{\Gamma\left(\frac{1}{2(d-1)}\right)}\right)^{d-1}\right].
\end{split}
\end{align}
In what follows we will use this $\Delta S$ notation in several parts of this paper.

Clearly the above expression does not hold for $d=2,4$. Indeed, for $d=4$ one finds
\begin{align}\label{stripHEE4}
\Delta S=\frac{L^{3}H^2}{16G_N}\left[\log\frac{\ell}{\epsilon}+\log\left(\frac{\Gamma\left(\frac{1}{6}\right)}{2^\frac{2}{3}\sqrt{\pi}\Gamma\left(\frac{2}{3}\right)}\right)-\frac{1}{3}\right] \alpha^2+\mathcal{O}\left(\alpha^4\right).
\end{align}
According to the above result in $d=4$, the momentum relaxation parameter $\alpha$ leads to appearance of subleading terms in the entropy expansion, including a logarithmic universal term in the HEE for infinite strip entangling region. The new universal term in this case is 
\bea\label{eq:str4d}
S_{\rm univ.}=\frac{L^{3}H^2}{16G_N}\alpha^2\log \frac{\ell}{\epsilon}.
\eea
The above perturbative analysis shows that for even $d$'s with $(d>3)$, there always exists a universal term at $\mathcal{O}\left(\alpha^{d-2}\right)$ of the perturbative expansion.
 
Finally for $d=2$ case we find
\begin{align}
\begin{split}
S&=\frac{L}{2G_N}\left[\log \frac{\ell}{\epsilon}+\frac{\alpha^2 \ell^2}{72}\left(2-\frac{3}{2}\log \alpha\ell\right)\right]+\mathcal{O}\left(\alpha^4\right).
\end{split}
\end{align}
Here the interesting point is that since the universal term of entanglement entropy coincides with the leading divergence, which is fixed by the UV structure of the theory, it does not get momentum relaxation corrections. As a matter of fact a non-critical (massive) contribution, which was first introduced in the celebrated work by Calabrese and Cardy \cite{Calabrese:2004eu}, has appeared as the leading momentum relaxation corrections.  

\subsubsection*{(ii) Case $T=0$ Mixed State}
Now we consider the extremal black-brane solution which we have previously introduced in \eqref{eq:PolyExt} and \eqref{eq:PolyExt2}. In this case the entanglement entropy for strip entangling region in the $\rho_h\to\infty$ limit is given by
\be
\Delta S=\frac{L^{d-1}H^{d-2}}{4G_N}\frac{d}{(d-2)(d-4)}\left[\frac{1}{\epsilon ^{d-4}}-\frac{3\sqrt{\pi}}{(d+2)\ell^{d-4}}\left(\frac{2\sqrt{\pi}\Gamma\left(\frac{d}{2(d-1)}\right)}{\Gamma\left(\frac{1}{2(d-1)}\right)}\right)^{d-4}\frac{\Gamma\left(\frac{3d}{2(d-1)}\right)}{\Gamma\left(\frac{2d+1}{2(d-1)}\right)}\right]\frac{1}{\rho_h^2}.
\ee
For the case of $d=4$ one finds
\be
\Delta S=\frac{L^3H^2}{2G_N}\left[\log\frac{\ell}{\epsilon}+\log\left(\frac{\Gamma \left(\frac{1}{6}\right)}{2^{2/3}\sqrt{\pi}\Gamma \left(\frac{2}{3}\right)}\right)\right]\frac{1}{\rho_h^2},
\ee
which again gives a correction to the universal term as in the non-extremal case reported previously in \eqref{eq:str4d}. 

For the case of $d=2$, the extremal geometry is given by $\alpha^2=\frac{4}{\rho_h^2}$.
The entanglement entropy in the $\rho_h\to\infty$ limit 
in terms of the momentum relaxation parameter is given by
\be\label{eq:PolyNC}
S=\frac{L}{2G_N}\left[\log \frac{2\ell}{\epsilon }-\frac{\alpha^2\ell^2}{4}\left(\frac{1}{3}\log \left(16\alpha\ell\right)-\frac{31}{18}\right)\right],
\ee
which again in this case a non-critical like correction appeared in entanglement entropy.

\subsubsection*{(iii) Case large entangling region}
Considering large entangling region limit, i.e., $\ell\gg \rho_h$, the main contribution to the area of the minimal surface comes from the limit where it is extended all the way to the horizon, such that $\rho_t\sim \rho_h$ (see \cite{LARGE} for related analysis). In this limit by defining $\rho=\rho_t \xi$, one finds
\be
\frac{\ell}{2}\approx\rho_h \int_0^{1} \frac{\xi^{d-1}d\xi}{
\sqrt{f(\xi)\left(1-\xi^{2(d-1)}\right)}},\;\;\;\;\;\;\;\;\;
\mathcal{A}\approx\frac{2L^{d-1}H^{d-2}}{\rho_h^{d-2}}\int_{\frac{\epsilon}{\rho_h}}^{1} \frac{d\xi}{\xi^{d-1}
\sqrt{f(\xi) \left(1-\xi^{2(d-1)}\right)}}.
\ee
Beside the UV divergent term in $\mathcal{A}$, the main contribution in the above integrals comes from the upper limit $\xi\to1$. Around this point we have
\be
\mathcal{A}\approx\frac{2L^{d-1}H^{d-2}}{\rho_h^{d-2} }\left(
 \int_0^{1} \frac{\xi^{d-1}d\xi}{
\sqrt{f(\xi)\left(1-\xi^{2(d-1)}\right)}}+
\int_{\frac{\epsilon}{\rho_h}}^{1} d\xi\; \frac{\sqrt{1-\xi^{2(d-1)}}}{\xi^{d-1}
\sqrt{f(\xi)}}\right).
\ee
Clearly the first term in the above expression is divergent as $\xi\to1$ while the second
one remains finite in this limit. Combining the above equation with the expression for $\ell$ one finds
\be
\mathcal{A}\approx\frac{L^{d-1}H^{d-2}}{\rho_h^{d-1} }\;\ell
 +\frac{2L^{d-1}H^{d-2}}{\rho_h^{d-2} }
\int_{\frac{\epsilon}{\rho_h}}^{1} d\xi\; \frac{\sqrt{1-\xi^{2(d-1)}}}{\xi^{d-1}
\sqrt{f(\xi) }}.
\ee
Here one should extract the UV divergent part of the second integral. Using
\be
f(\xi)=1-\xi^d-\frac{\alpha^2\rho_h^2}{2(d-2)}(\xi^2+\xi^d),
\ee
and 
\be
\int_{\frac{\epsilon}{\rho_h}}^{1} d\xi\; \frac{\sqrt{1-\xi^{2(d-1)}}}{\xi^{d-1}
\sqrt{f(\xi)}}=\frac{\rho_h^{d-2}}{(d-2)\epsilon^{d-2}}-c_d+\alpha^2\rho_h^2\left(A_d(\epsilon)+c_d'\right)+\mathcal{O}\left(\alpha^4\right),
\ee
where
\be
A_3(\epsilon)=0,\;\;\;A_4(\epsilon)=\frac{1}{8}\log \frac{\rho_h}{\epsilon},\;\;\;A_5(\epsilon)=\frac{\rho_h}{12\epsilon},\;\;\;\cdots,
\ee
and $c_d$ and $c_d'$ are numerical factors, one can easily find that the entanglement entropy in this limit reads
\be\label{SBH1}
S\approx\frac{2L^{d-1}H^{d-2}\rho_h^{d-2}}{4G_N }\left[\frac{1}{(d-2)\epsilon^{d-2}}+\frac{\ell}{2\rho_h^{d-1}}
-\;\frac{c_d}{\rho_h^{d-2}}+\frac{\alpha^2}{\rho_h^{d-4}}\left(A_d(\epsilon)+c_d'\right)\right]+\mathcal{O}(\alpha^4).
\ee

The second term in the above expression is the thermal entropy which is proportional to the volume. Moreover, the remaining terms are proportional to the area of the entangling region. There are no $\mathcal{O}\left(\alpha^2\right)$ corrections to the thermal entropy. Also one can easily redo the same calculations for the extremal case and find similar results.

\subsection{Spherical and Cylindrical Entangling Regions}

As we have mentioned previously, the momentum relaxation parameter may affect the universal part of HEE. Here at this stage we are mainly interested in spherical and cylindrical regions. It is well known that in even-dimensional CFTs there exists a logarithmic universal term in the entanglement entropy expansion \cite{Ryu:2006ef}. The reason is that these two entangling regions are shown to capture the `a'-type and `c'-type anomalies in 4-dimensional CFTs \cite{Solodukhin:2008dh, Hung:2011}. The `a'-type anomaly is the coefficient of the four-dimensional Euler density and the`c'-type anomaly is the coefficient of the Weyl squared tensor in the trace of stress tensor.\footnote{In the case of infinite strip entangling region it is believed that the cut-off independent part of entanglement entropy is a complicated function of the anomaly coefficients in 4-dimensional CFTs \cite{Singh:2012ala, Myers:2012ed}.} 

Lets first consider the momentum relaxation correction to the universal part of spherical entangling region in $d=4$. The corresponding hypersurface in the bulk can be parametrized as $r=r(\rho)$ and the induced metric on the hypersurface is given by
\be
ds^2_{\rm ind.}=\frac{L^2}{\rho^2}\left[\left(\frac{1}{f}+r'^2\right)d\rho^2+r^2 d\Omega_{d-2}^2\right],
\ee
where prime denotes the derivative with respect to $\rho$. In this case the area can be computed as 
\bea\label{sphereHEE}
\mathcal{A}=L^{d-1}\Omega_{d-2}\int d\rho \frac{r^{d-2}}{\rho^{d-1}}\sqrt{\frac{1}{f}+r'^2}.
\eea
In this case we are again interested in first non-trivial $\alpha$ corrections to the profile of the minimal surface in the bulk.
The final result for the HEE is given by
\bea
S=\frac{\pi L^3}{2G_N}\left[\frac{\ell^2}{\epsilon^2}-\frac{1}{2}-\log\frac{2\ell}{\epsilon}+\frac{\alpha^2\ell^2}{4}\left(\log\frac{2\ell}{\epsilon}-\frac{4}{3}\right)\right]+\mathcal{O}\left(\alpha^4\right).
\eea
It is well established that for theories dual to Einstein gravity, $a=\frac{\pi L^3}{8G_N}$ and for higher curvature gravity theories this coefficient is modified due to stringy corrections. This result shows that due to the presence of the momentum relaxation parameter this universal term is also modified as follows
\bea
S_{\rm univ.}\sim -4\mathfrak{a}\log\frac{2\ell}{\epsilon},\;\;\;\mathfrak{a}=a\left(1-\frac{\alpha^2 \ell^2}{4}\right).
\eea
The above result could be interpreted as the change of the central charge as the theory is slightly deformed by the marginal deformations corresponding to momentum relaxation. Since the correction decreases the value of the central charge at the fixed point it is in agreement with `a'-theorem.

Now let us consider a cylindrical entangling region. In this case the corresponding hypersurface can be parametrized as $r=r(\rho)$ and the induced metric on the hypersurface is given by
\be
ds^2_{\rm ind.}=\frac{L^2}{\rho^2}\left[\left(\frac{1}{f}+r'^2\right)d\rho^2+du^2+r^2 d\Omega_{d-3}^2\right],
\ee
where prime denotes the derivative with respect to $\rho$. Using the above metric the area can be written as
\bea\label{cylindrHEE}
\mathcal{A}=L^{d-1}H\Omega_{d-3}\int d\rho \frac{r^{d-3}}{\rho^{d-1}}\sqrt{\frac{1}{f}+r'^2}.
\eea
The equation of motion can not be solved analytically even considering perturbations around $\alpha=0$. As long as we are interested in logarithmic universal terms, the near boundary behavior of the minimal surface is enough to read this universal contribution. Following \cite{Hung:2011} we expand the profile of the hypersurface near the boundary, $\rho=0$. Since the equation of motion of $r$ is even under $\rho\rightarrow - \rho$, only even powers of $\rho$ appears. 
Explicit computations for $d=4$ leads to
\bea
r(\rho)=\ell-\frac{\rho^2}{4\ell}+\cdots.
\eea 
Finally one can find the universal contribution of HEE in $d=4$ as
\bea
S_{\rm univ.}=\frac{\pi L^3H}{2 G_N}\frac{-1+\alpha^2 \ell^2}{8\ell}\log\frac{\ell}{\epsilon}.
\eea
The entanglement entropy of a cylinderical entangling region in a 4-dimensional CFT has a universal term as $S_{\rm univ.}\sim -\frac{c}{2}\frac{H}{\ell}\log\frac{2\ell}{\epsilon}$ where $c$ is the coefficient of the Weyl squared tensor in the trace anomaly expression \cite{Solodukhin:2008dh, Hung:2011}. In four dimensions for theories dual to Einstein gravity we have $c=\frac{\pi L^3}{8G_N}$ and stringy corrections modifies it \cite{Hung:2011}. So the above result shows that this universal term changes as 
\bea
S_{\rm univ.}\sim -\frac{\mathfrak{c}}{2}\frac{H}{\ell}\log\frac{\ell}{\epsilon},\;\;\;\mathfrak{c}=c\left(1-\alpha^2 \ell^2\right).
\eea
One may interpret this corrected universal term as the corrected `c'-type central charge of the dual theory which interestingly has decreased along the flow triggered by the momentum relaxation marginal deformation.

\subsection{Momentum Relaxation and Hyperscaling Violation}

It is well known in the literature that if the leading divergence of entanglement entropy is a logarithmic term, the corresponding system has a Fermi surface \cite{FS}. This was realized in holography via hyperscaling violating geometries in \cite{Ogawa:2011bz} and \cite{Huijse:2011ef} (see also \cite{Takayanagi:2013ywa} for a review).\footnote{This is based on a generalization of the RT proposal to non-AAds solutions of Einstein gravity.} To be more precise, in terms of the parameters of this geometry which we have previously introduced a version with momentum relaxation in \eqref{hsmetric}, for the choice of $\theta=d-1$ this geometry is believed to be dual to phases of matter with `hidden' Fermi surface of `fractionalized' degrees of freedom (for details see \cite{Huijse:2011ef}).

It would be interesting to study the effect of momentum relaxation deformation on the formation of this kind of hidden Fermi surface in holography. To do so we consider \eqref{hsmetric} for a $d+2$ dimensional bulk theory and study entanglement entropy for the dual state of this geometry. 
Considering a strip entangling region leads to the following area functional
\be
\mathcal{A}=2H^{d-1}\int_\epsilon^{\rho_t}\rho^{\theta-d}\sqrt{{x'}^2(\rho)+\frac{1}{f(\rho)}}d\rho.
\ee
The equation for the minimal hypersurface can be solved using a conserved quantity in the above action. This leads to
\be
{x'}(\rho)=\pm\frac{\rho^{d-\theta}}{\sqrt{f(\rho)\left(\rho_t^{2(d-\theta)}-\rho^{2(d-\theta)}\right)}},
\ee
and the HEE is given by
\be\label{eq:intgrandSHS}
S=\frac{H^{d-1}}{2G_N}\int_\epsilon^{\rho_t}\frac{\rho_t^{d-\theta}}{\rho^{d-\theta}}\frac{d\rho}{\sqrt{f(\rho)\left(\rho_t^{2(d-\theta)}-\rho^{2(d-\theta)}\right)}}.
\ee
Setting $m_0=0$ the leading contribution to the HEE due to the momentum relaxation parameter is given by

\begin{align}\label{eq:HEEHS}
\begin{split}
\Delta S=\frac{H^{d-1}}{4G_N}&\frac{d}{\left(d^2-d (\theta +2 z+1)+2 \theta \right)}\times\\&
\left[\frac{1}{\epsilon ^{\frac{2 \theta }{d}+d-\theta-2 z-1}}-\sqrt{\pi }\frac{\Gamma \left(\frac{1}{2}+\frac{2 (d z-\theta )+d}{2 d(d-\theta)}\right) }{ \Gamma \left(\frac{2(z d-\theta)+d}{2 d(d-\theta)}\right)}\left(\frac{\mathcal{Q}}{\ell}\right)^{\frac{2 \theta }{d}+d-\theta-2 z-1}\right]\alpha^2+\mathcal{O}\left(\alpha^4\right),
\end{split}
\end{align}
where
\be
\mathcal{Q}\equiv\frac{2\sqrt{\pi}\Gamma \left(\frac{d-\theta +1}{2 d-2 \theta }\right)}{\Gamma \left(\frac{1}{2 d-2 \theta }\right)}.
\ee
To see whether this model is going to have leading logarithmic divergence in entanglement entropy, that is formation of a hidden Fermi surface, we just have to look at the expansion of the integrand of entanglement entropy in \eqref{eq:intgrandSHS}. The integrand in momentum relaxation parameter expansion is given by
\be
\frac{\left(\frac{\rho_t}{\rho}\right)^{d-\theta }}{\sqrt{\rho_t^{2 d-2 \theta }-\rho^{2 d-2 \theta }}}+\frac{\alpha ^2 \left(\frac{\rho_t}{\rho}\right)^{d-\theta } \rho^{2 z-\frac{2 \theta }{d}}}{2 \sqrt{\rho_t^{2 d-2 \theta }-\rho^{2 d-2 \theta }}}+\mathcal{O}\left(\alpha^4\right).
\ee
As we mentioned above, for $d-\theta=1$ the zeroth order gives a logarithmic divergence in EE. Now we would like to see whether the first non-trivial $\alpha$ correction to this expression may contribute at this order. This would happen for
$$d-\theta-2z+2\frac{\theta}{d}=1,$$
for which if we apply $d-\theta=1$, it gives
$$z=\frac{\theta}{\theta+1}.$$
One can easily check that the above condition violates the null energy conditions of the background, which can also be found from reality conditions on the dilaton and gauge fields of the corresponding solution in \eqref{hsmetric} as
\begin{align}
(z-1)(z+d-\theta)&>0,\\
(d-\theta)(z-1-\theta/d)&>0.
\end{align}

We have shown that the Fermi surface does not get correction from momentum relaxation parameter but if the dual state does not admit a Fermi surface at $\alpha=0$, that is $d-\theta\neq 1$, we can easily find windows in the parameter space of $(\theta,d,z)$ where a logarithmic correction may appear in the expression of entanglement entropy with
$$d-\theta-2z+2\frac{\theta}{d}=1.$$

It is worth to note that another interesting feature of the above result \eqref{eq:HEEHS} is the appearance of dynamical exponent, i.e., $z$, in the expression for HEE. To our knowledge this was not previously seen in any static state of theories dual to Einstein gravity. Although in the case of non-static states the dynamical exponent may appear in the entanglement entropy (see \cite{Alishahfonda}).\footnote{Also in the case of higher derivative gravity theories, appearance of the dynamical critical exponent in the entanglement entropy was previously reported in \cite{Hosseini:2015gua, Basanisi:2016hsh}.}\footnote{See also \cite{Ling:2016ien} for related studies in hyperscaling violating backgrounds with momentum relaxation.}

\section{Other Holographic Entanglement Measures}\label{sec:other}

In this section we will study other holographic measures of entanglement such as mutual information and information metric.

\subsection{Mutual Information}
Entanglement entropy is in general a divergent quantity which does not capture much from the field theory in hand. For instance for a single interval entangling region in a two dimensional CFT it only depends on the central charge of the theory. In order to have a finite measure which contains more information about the field content of the theory one can employ other quantities such as mutual information. Mutual information quantifies the extent which the degrees of freedom of two subsystems are correlated with each other and is defined by
\bea
I(A_1, A_2)=S_{A_1}+S_{A_2}-S_{A_1\cup A_2},
\eea
where $S_{A_1\cup A_2}$ is the entanglement entropy for the union of two subsystems. Using  subadditivity property of the entanglement entropy it is obvious that mutual information is always positive. Although for disjoint regions the divergent terms appear in the expression for mutual information cancel each other, but it becomes divergent when these regions share boundaries \cite{{Headrick:2010zt},{Mozaffar:2015xue}}. 

For holographic CFTs mutual information can be computed using the RT prescription. Actually due to the competition between two different configurations corresponding to $S_{A_1\cup A_2}$, it was shown that holographic mutual information exhibits a phase transition\cite{Headrick:2010zt}. The location of the critical point depends on the ratio of the length of the entangling regions to their separation. For large entangling regions with small separation, the holographic mutual information is finite and it vanishes in the opposite limit where the correlation between these regions becomes negligible. 

\begin{figure}
\begin{center}
\includegraphics[scale=0.45]{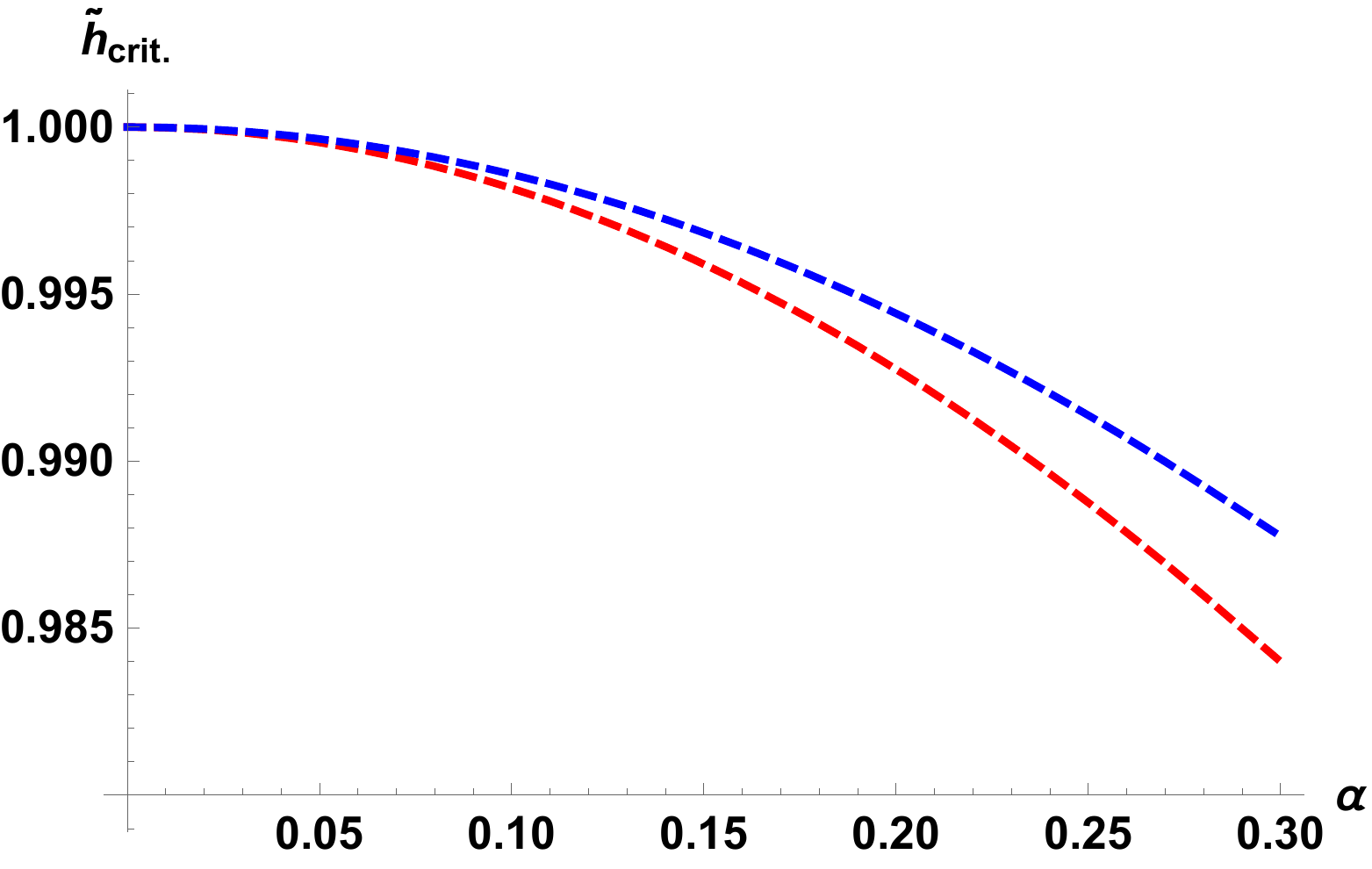}
\end{center}
\caption{Location of the critical points as a function of $\alpha$ for $d=3$ (red) and $d=4$ (blue) with $\ell=1$ in polynomial  model. In each case the holographic mutual information is finite below the transition curve and vanishes when we cross it.}
\label{fig:mutual1}
\end{figure}

In order to investigate the effects of momentum relaxation on the mutual information, we compute this quantity for two disjoint strips where their lengths and separation are given by $\ell_1$, $\ell_2$ and $h$ respectively. In this case we have
\bea\label{SAUB}
S_{A_1\cup A_2}=\Bigg\{ \begin{array}{rcl}
&S(\ell_1+\ell_2+h)+S(h)&\,\,\,h\ll \ell_1, \ell_2,\\
&S(\ell_1)+S(\ell_2)&\,\,\,h\gg \ell_1, \ell_2,
\end{array}\,\,
\eea
and the mutual information becomes
\bea\label{HMI}
I(A_1, A_2)=\Bigg\{ \begin{array}{rcl}
&S(\ell_1)+S(\ell_2)-S(\ell_1+\ell_2+h)-S(h)&\,\,\,h\ll \ell_1, \ell_2,\\
&0&\,\,\,h\gg \ell_1, \ell_2.
\end{array}\,\,
\eea
In order to simplify the computations, we will set $\ell_1=\ell_2$ in what follows. Using the expression of the HEE for a strip entangling region, i.e., \eqref{stripHEE}, the mutual information becomes\footnote{In the following we neglect an overall factor of $\frac{L^{d-1}H^{d-2}}{2(d-2)G_N}$ which is positive and does not change our results about the phase transitions and location of the critical points.}
\bea
\Delta I=c_1 \left(\frac{2}{\ell^{d-4}}-\frac{1}{(2\ell+h)^{d-4}}-\frac{1}{h^{d-4}}\right)\alpha^2,\;\;\;c_1=\frac{2^{d-7}\sqrt{\pi}}{3}\left(\frac{\sqrt{\pi}\Gamma\left(\frac{d}{2(d-1)}\right)}{\Gamma\left(\frac{1}{2(d-1)}\right)}\right)^{d-4}\frac{\Gamma\left(\frac{4-d}{2(d-1)}\right)}{\Gamma\left(\frac{3}{2(d-1)}\right)},
\eea
where we have subtracted $\alpha=0$ contribution, i.e., $\Delta I=I-I_0$, and
\begin{align}
\begin{split}
I_0=-c_0\left(\frac{2}{\ell^{d-2}}-\frac{1}{(2\ell+h)^{d-2}}-\frac{1}{h^{d-2}}\right),\;\;
c_0=2^{d-2}\left(\frac{\sqrt{\pi}\Gamma\left(\frac{d}{2(d-1)}\right)}{\Gamma\left(\frac{1}{2(d-1)}\right)}\right)^{d-1}.
\end{split}
\end{align}
Also for $d=4$ using \eqref{stripHEE4} one finds
\bea
\Delta I=\frac{\alpha^2}{4}\log\frac{\ell^2}{h(2\ell+h)}.
\eea
Clearly the position of the critical point where the holographic mutual information vanishes depends on the momentum relaxation parameter $\alpha$. For example, in the case of $d=3$ for small $\alpha$ one finds 
\bea
h_{\rm crit.}=\frac{1}{2} \left(\sqrt{5}-1\right) \ell-\frac{\alpha ^2 \ell^3 \Gamma \left(\frac{1}{4}\right)^4}{96 \sqrt{5} \pi  \Gamma \left(\frac{3}{4}\right)^4}+\mathcal{O}(\alpha^4).
\eea
We have summarized the results for $d=3$ and $d=4$ with $\ell=1$ in Fig.\ref{fig:mutual1}.
In order to compare the results more clearly, we plot $\tilde{h}_{\rm crit.}=\frac{{h}_{\rm crit.}}{{h}^{\alpha=0}_{\rm crit.}}$ as a function of $\alpha$. In this figure the corresponding holographic mutual information is finite below each curve and it vanishes above them. This shows that in theories with momentum relaxation the phase transition of mutual information happens at smaller separation between the spatial subsystems comparing to translational invariant states. In other words it means that mutual correlation between subsystems is a decreasing function of momentum relaxation parameter. 

This analysis can be generalized to other holographic information measures to study how momentum relaxation affects their behaviour. The simplest example is the holographic tripartite information which is defined as follows
\bea\label{tripartite}
I^{[3]}(A_1, A_2, A_3)=I(A_1,A_2)+I(A_1,A_3)-I(A_1,A_2\cup A_3).
\eea
In \cite{Hayden:2011ag} it was proved that this quantity is always negative which means that the holographic mutual information is monogamous. Note that in a general QFT the tripartite information can be positive, negative, or zero and it seems that this monogamy property is a necessary condition for a QFT to admit Einstein dual gravity. Also it was shown that in specific situations the holographic $n$-partite information which is a generalization of \eqref{tripartite} to systems consisting of $n$ subsystems has a definite sign, i.e., it is positive (negative) for even (odd) $n$\cite{npar}. Actually the presence of momentum relaxation does not change these arguments. The basic assumptions leading to these behaviors are the minimality and homology conditions of the RT prescription. These two are supposed to hold in solutions dual to momentum dissipation as long as they are asymptotically AdS geometries as solutions of Einstein gravity with minimally coupled matter fields.

\subsection{Information Metric}

Fisher information metric (sometimes called Bures metric, information metric or even quantum fidelity) is a measure to quantify how much two different states are different. It measures the distance between states in the states space. This quantity is defined for states which are infinitesimally apart from the fidelity expansion
\be
\mathcal{F}(\alpha_0,\alpha_0+\delta\alpha)\equiv\left|\langle\psi(\alpha_0+\delta\alpha)|\psi(\alpha_0)\rangle\right|=1-G_{\alpha\alpha}\left(\delta\alpha\right)^2 +\mathcal{O}\left(\left(\delta\alpha\right)^3\right),
\ee
where $G_{\alpha\alpha}$ is defined as the information metric or fidelity susceptibility. This measure has several applications including a useful tool to understand quantum critical phenomena and quantum phase transitions \cite{fidelity0} (see also e.g. \cite{fidelity} for a review).

Recently a proposal for holographic information metric has appeared for states of a $(d+1)$-dimensional CFT which are separated due to a marginal deformation \cite{MIyaji:2015mia} (see \cite{Bak:2015jxd} for a concrete generalization of this proposal). This proposal is supposed to work for states which their dual geometry are static solutions of Einstein gravity. The proposal simply says that the information metric can be calculated holographically via
\be
G_{\alpha\alpha}^{(d+1)}=n_d\frac{\mathrm{Vol}(\Sigma_{\mathrm{max}})}{L^{d+1}},
\ee
where $n_d$ is a numerical factor of $\mathcal{O}(1)$, $\Sigma_{\mathrm{max}}$ is a time-slice (co-dimension one) with maximum volume of the geometry which ends on the boundaries of the geometry and $L$ is the AdS radius. 

In this subsection we compute information metric between two states at leading order of mass and momentum relaxation parameter corrections.
Explicit computations for 2d CFT leads to 
\begin{align}
\begin{split}
G^{(2)}_{\alpha\alpha} &=\frac{V_2}{2L} \left(\frac{1}{\epsilon ^2}-\frac{1}{\rho_h^2}\right)-{\frac{V_2\alpha ^2}{4L} \log\frac{\epsilon}{\rho_h}}.
\end{split}
\end{align}
where for higher dimensional CFTs $G^d_{\alpha\alpha}$ is given by
\bea
G^{(d+1)}_{\alpha\alpha}=\frac{V_{d-1}}{(d-1)L}\left(\frac{1}{\epsilon^{d-1}}-\frac{1}{\rho_h^{d-1}}\right)+\frac{V_{d-1}\alpha^2}{4(d-2)(d-3)L}\left(\frac{1}{\epsilon^{d-3}}-\frac{1}{\rho_h^{d-3}}\right).
\eea
The above expressions can be viewed two-fold. In order to find the fidelity susceptibility between the conformal vacuum (dual to the pure AdS geometry) and the massless non-conformal vacuum one may take $\rho_h\to\infty$ limit of these expressions. On the other hand the whole expressions give the fidelity susceptibility between massless non-conformal vacuum and a massive deformation of that.\footnote{One may also consider a more recent proposal known as `holographic complexity' for holographic calculation of `reduced fidelity susceptibility' \cite{Alishahiha:2015rta}.}

\section{Geometric Entropy and Confinement/deconfinement Phase Transition}\label{sec:geo}

An interesting property of holographic entanglement entropy is probing confinement/deconfinement phase transitions \cite{Nishioka:2006gr, Klebanov:2007ws} (see also \cite{Nishioka:2009un} for a review). This property is captured by solitonic solutions which are obtained after a double Wick rotation. Such a rotation in our case \eqref{staticmetric}, leads to
\begin{align}\label{solitonicmetric1}
\begin{split}
ds^2&=\frac{L^2}{\rho^2}\left[f(\rho)dt^2+\frac{d\rho^2}{f(\rho)}-dx_1^2+dx_{d-2}^2\right].
\end{split}
\end{align}
In this section we are interested in studying the effect of momentum relaxation on these kind of phase transitions. We restrict ourselves to the case of $d=3$ with $m_0=0$. 
Note that here the $t$ direction is compacted. Imposing the condition that there is no conical singularity at the horizon $\rho_h=\frac{\sqrt{2}}{\alpha}$, fixes the radius of $t$ to be $\beta=\frac{2\sqrt{2}\pi}{L \alpha}$.
In this case there are two types of RT surfaces which contribute to the HEE, and they are usually referred to as the connected and disconnected
RT surfaces. There is a critical value $\ell_c$ for the width $\ell$ of the strip entangling region, for which $\ell<\ell_c$ the area of the connected RT surface is minimal while for $\ell>\ell_c$ the disconnected RT surface is minimal.
\begin{figure}
	\begin{center}
		\includegraphics[scale=0.45]{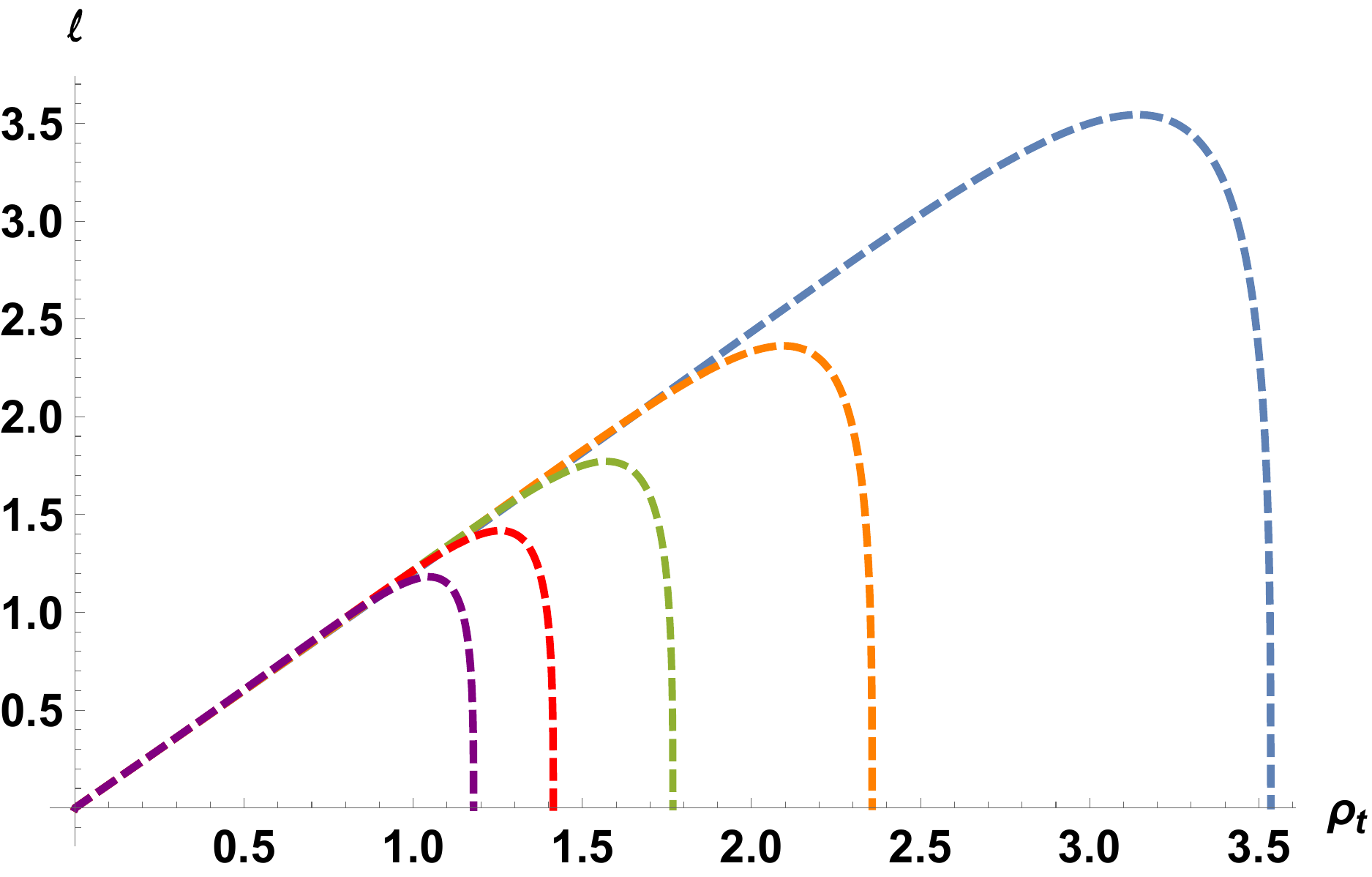}
	\end{center}
	\caption{The width of the strip $\ell$ as a function of the turning point $\rho_t$ for $\alpha=0.4, 0.6, \cdots$ from right to left. Note that for each $\ell$ there are always two solutions (two local minima of the area functional). However only one of them minimizes the area functional, and is called the physical connected configuration.}
\label{fig:soliton-l-rhot}
\end{figure}
\subsubsection*{Connected RT surfaces}
The connected RT surface is a smooth surface which starts from one boundary of strip and ends on the other boundary.
This surface is parametrized by $x_1=x_1(\rho)$, thus the induced metric on it is given by
\be
ds^2_{\rm ind.}=\frac{L^2}{\rho^2}\left[f(\rho) dt^2+\left({x_1'}^2+\frac{1}{f(\rho)}\right)d\rho^2\right].
\ee
The area functional for the connected RT surfaces is given by
\bea
\mathcal{A}^{con}=\int dtdx_1 \frac{L^2}{\rho^2}\sqrt{\rho'+f(\rho)}.
\eea
The Hamiltonian $\mathcal{H}$ conjugate to $x_1$ is given by
\bea
\mathcal{H}=\frac{f(\rho)}{\rho^2\sqrt{{\rho'}^2+f(\rho)}}=\frac{\sqrt{f(\rho_t)}}{\rho_t^2},
\eea 
and is a conserved quantity. Using the hamiltonian $\mathcal{H}$ one can find
\bea
\rho'(x_1)=\pm\frac{\sqrt{\left(\rho_t^2-\rho^2\right) \left(-2+\alpha^2 \rho^2\right) \left(2\rho_t^2 +\rho^2\left(2-\alpha^2\rho_t^2\right)\right)}}{\rho^2\sqrt{2\alpha^2\rho_t^2-4}},
\eea
which $\rho_t$ is related to the boundary data via
\begin{align}\label{soliton-l}
\ell=2\int_0^{\rho_t} d\rho\frac{1}{\rho'}.
\end{align}
In Fig. \eqref{fig:soliton-l-rhot},  $\ell$ is plotted as a function of $\rho_t$. For the connected configuration there are always
two solutions to the e.o.m. of $\rho(x_1)$ : the solution which gives the smallest area is called
`physical', and the other one is called `unphysical' RT surface in the literature. The entropy given
by the physical connected RT surface is
\begin{align}\label{soliton-HEE-con}
S^{\mathrm{con.}}=\frac{\beta L^2}{2 G_N} \int_\epsilon^{\rho_t} d\rho \left(\frac{\rho_t}{\rho}\right)^2 \sqrt{\frac{\left(2-\alpha^2 \rho^2\right)}{\left(\rho_t^2-\rho^2\right) \left(2\rho_t^2+\rho^2 \left(2-\alpha^2 \rho_t^2\right)\right)}}.
\end{align}
\begin{figure}
	\begin{center}
		\includegraphics[scale=0.45]{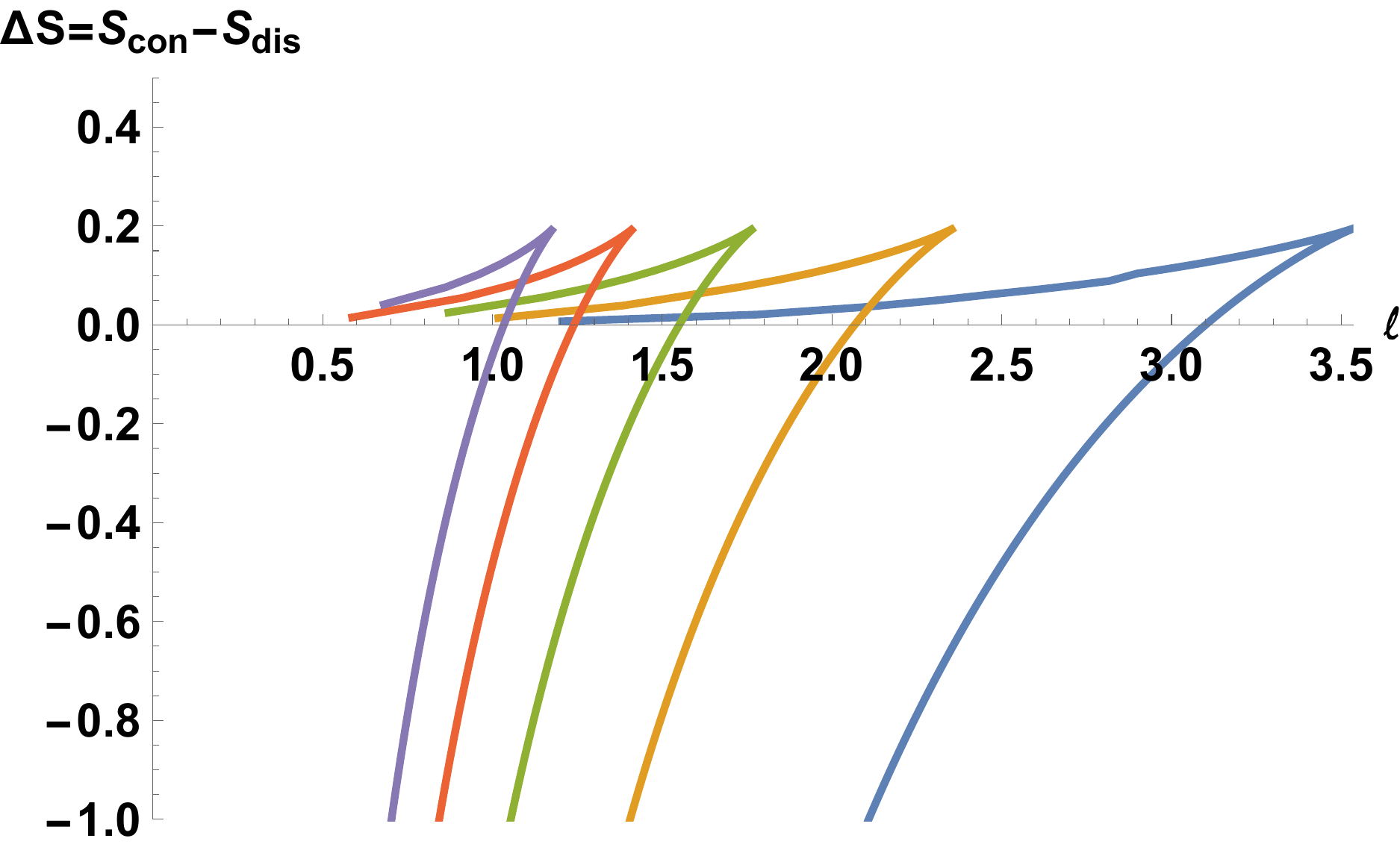}
	\end{center}
	\caption{The subtracted holographic entanglement entropy $\Delta S=S^{\mathrm{con.}}-S^{\mathrm{dis.}}$ as a function of the strip width $\ell$ for $\alpha=0.4, 0.6, \cdots$ from right to left. For each $\alpha$ there is a curve which has two branches: upper and lower $\Delta S=0$. The upper branch is what is usually called unphysical while the lower branch is the global minimum of the area functional and shows the physical connected RT surface in the HEE. Here we set $\epsilon=0.001$, $L=G_N=1$}
\label{fig:soliton-deltaS}
\end{figure}
\subsubsection*{Disconnected RT surfaces}
\begin{figure}
	\begin{center}
		\includegraphics[scale=0.45]{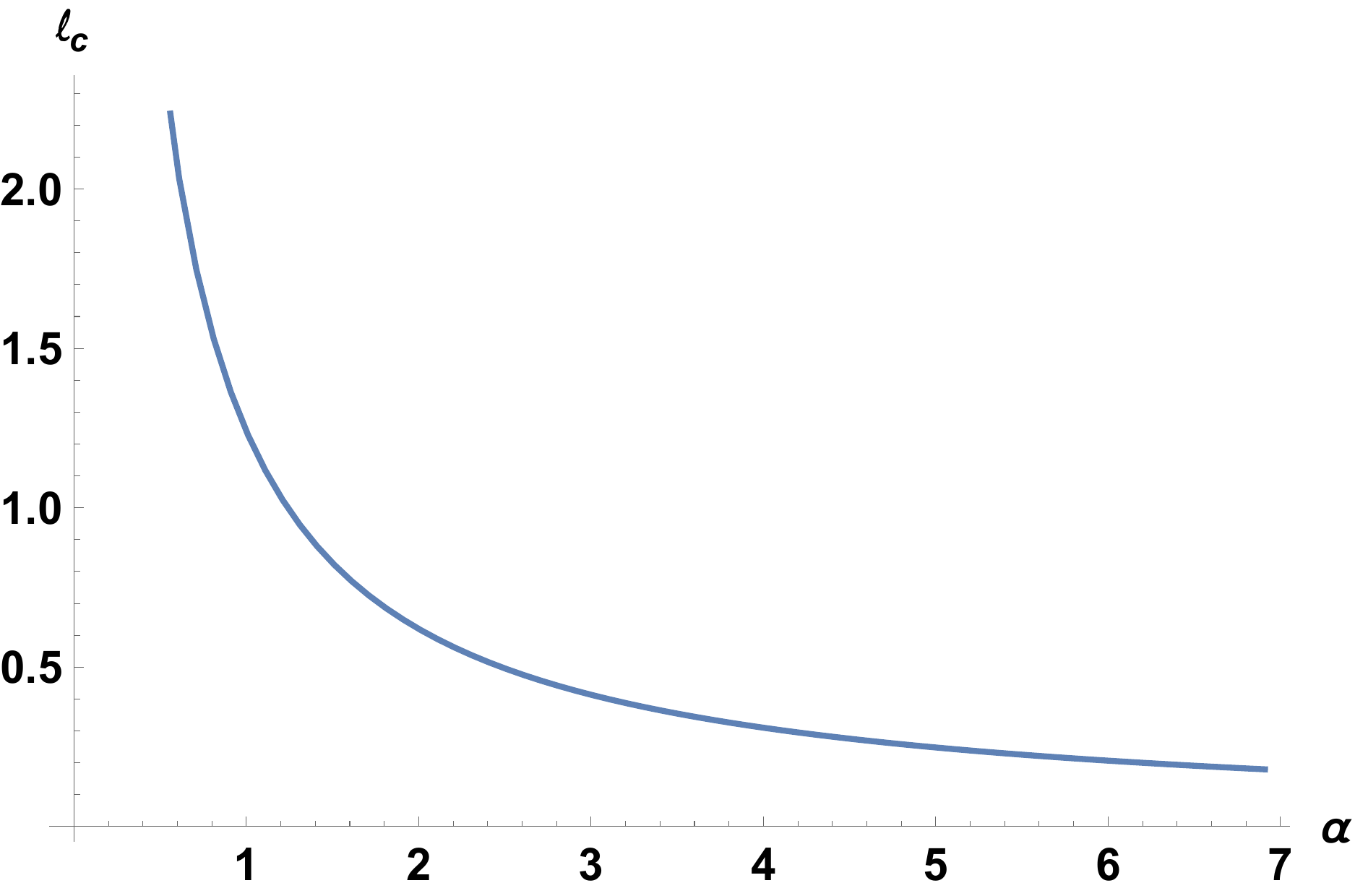}
	\end{center}
	\caption{Critical length $l_c$ as a function of $\alpha$. By increasing the parameter $\alpha$, the critical length decreases.}
\label{fig:soliton-lc}
\end{figure}

The disconnected RT surface is a union of two disconnected parts each starting from one boundary towards the horizon at $\rho_h$. In this case $x_1$ is independent of the radial coordinate $\rho$, in contrast to the connected case. The induced metric is given by
\be
ds^2_{\rm ind.}=\frac{L^2}{\rho^2}\left[f(\rho) dt^2+\frac{d\rho^2}{f(\rho)}\right],
\ee
and the HEE is given by
\bea\label{soliton-HEE-dis}
\begin{split}
S^{\mathrm{dis.}}
=\frac{\beta L^2}{2 G_N} \left(\frac{1}{\epsilon}-\frac{1}{\rho_h} \right).
\end{split}
\eea 
We are interested in the difference between the contributions of these two type of RT surfaces. Therefore, we consider the subtracted EE, $\Delta S=S^{\mathrm{con.}}-S^{\mathrm{dis.}}$ which is a UV finite quantity.
In Fig. \eqref{fig:soliton-deltaS} we have plotted $\Delta S$ as a functions of $\ell$.
As can be seen from Fig. \eqref{fig:soliton-deltaS} by increasing the momentum relaxation parameter $\alpha$,  the critical length $\ell_c$ decreases. We have also plotted $\ell_c$ as a function of $\alpha$ in Fig. \eqref{fig:soliton-lc}.

\section{Wilson Loop}\label{sec:wilson}

In this section we will compute another nonlocal probe which is a rectangular Wilson loop in the geometry given by \eqref{staticmetric}. Using the expectation value of this quantity one can read off the effective potential between extrenal point like objects, e.g., a quark-antiquark pair. The prescription for calculating the expectation values of Wilson loop operators in the dual theory was proposed in \cite{Maldacena:1998im}.\footnote{See \cite{Giataganas:2012zy} for studying the Wilson loop in the dual theory of anisotropic axionic backgrounds.} According to this proposal the corresponding expectation value is equal to the area of a worldsheet whose boundary is the loop located on the asymptotic boundary of the spacetime. The corresponding area for the string worldsheet is given by the Nambu-Goto action
\bea
I=\frac{1}{2\pi \alpha'}\int d\tau d\sigma \sqrt{h},
\eea
where $h$ is the induced metric on the worldsheet. Considering a static configuration with
\bea
\tau=t,\;\;\;\sigma=\rho,\;\;\;x_1=x(\rho),
\eea
one finds
\bea
I=\frac{L^2}{2\pi \alpha'}\int dt \frac{d\rho}{\rho^2}\sqrt{1+x'(\rho)^2f(\rho)}.
\eea
The action does not depend on $x$, hence by defining a constant of motion one finds
\bea
x'^2=\frac{\rho^4f(\rho_t)}{f(\rho)(f(\rho)\rho_t^4-f(\rho_t)\rho^4)},
\eea
where $\rho_t$ denotes the turning point. Using the above relation, the separation and also effective potential between the quark and antiquark can be found as follows
\begin{align}
\begin{split}
\ell&=2\int_0^{\rho_t}d\rho\rho^2\sqrt{\frac{f(\rho_t)}{f(\rho)(f(\rho)\rho_t^4-f(\rho_t)\rho^4)}},\\
V&=\frac{L^2\rho_t^2}{\pi\alpha'}\int_\epsilon^{\rho_t}\frac{d\rho}{\rho^2}\sqrt{\frac{f(\rho)}{f(\rho)\rho_t^4-f(\rho_t)\rho^4}}.
\end{split}
\end{align}
Since we are not able to perform the above integral analytically, we compute the first order correction due to the momentum relaxation parameter $\alpha$. This can be found as
\bea\label{deltaveff}
\Delta V=-\frac{L^2}{16\sqrt{\pi}(d-2)\alpha'}\frac{\Gamma({\frac{5}{4}})}{\Gamma({\frac{7}{4}})}\alpha^2\ell,
\eea
Similar to our previous notation we have defined $\Delta V=V-V^{(0)}$. The regularized part of the effective potential for the AdS vacuum is given by \cite{Maldacena:1998im}
\bea
V_{\rm reg.}^{(0)}=-\frac{2L^2}{\alpha'}\frac{\Gamma({\frac{3}{4}})^2}{\Gamma({\frac{1}{4}})^2}\frac{1}{\ell}.
\eea

According to \eqref{deltaveff} in theories with momentum dissipation the correction to the effective potential between point like external objects is linear and attractive. This result shows that the strength of the corresponding force between these objects is given by
\bea
F_0=-\frac{dV_0}{d\ell}=-\frac{2L^2}{\alpha'}\frac{\Gamma(\frac{3}{4})^2}{\Gamma(\frac{1}{4})^2}\frac{1}{\ell^2},
\eea 
which is an attractive force. On the other hand the momentum relaxation parameter leads to the following correction
\bea
\Delta F=-\frac{d \Delta V}{d \ell}=\frac{L^2}{96\sqrt{2}\pi^2 (d-2) \alpha'}\frac{\Gamma(\frac{1}{4})^3}{\Gamma(\frac{3}{4})} \alpha^2,
\eea 
which is repulsive and independent of the separation $\ell$. The shows that total force between the quark and the anti-quark decreases in presence of momentum dissipation.

\section{Discussions and Concluding Remarks}

In this section we would like to first summarize our results and continue with some complementary material of our study. We will  report the result of the main parts of our analysis for another similar type of momentum relaxation model known as square root model, and we will end with some comments mainly about massive deformations of the non-conformal vacuum and the first law of entanglement in theories with momentum relaxation.

\subsection*{Summary of Results}
\begin{itemize}
\item In the case of strip entangling region, it is well known from the very beginning of RT proposal that in generic $d$-dimensional (with $d>2$) holographic CFTs that there is no logarithmic universal term in the entanglement entropy expansion. Here we show that due to momentum relaxation effects, logarithmic universal terms may appear in the entanglement entropy expansion with respect to momentum relaxation parameter.

\item It is well-known that the universal terms of spherical entangling region capture the `a'-type and cylindrical entangling region capture the `c'-type central charges of 4-dimensional CFTs. Here we show that in presence of the marginal deformation of the CFT these universal terms get corrections from the momentum relaxation parameter (in agreement with a-theorem in case of spherical regions).

\item In the case of 2-dimensional CFTs, since the universal term is the leading divergence of entanglement entropy which is completely fixed from the UV structure of the theory, the marginal deformation does not affect the universal term and thus the central charge. In this situation entanglement entropy gets non-critical corrections due to momentum relaxation.

\item We have shown that increasing the distance between two subregions, in comparison with the conformal vacuum state,
the phase transition of holographic mutual information happens at smaller distance. This is because of the decrease of the correlation length in such states with momentum dissipation.

\item We have studied the phase transition captured by the double Wick-rotated geometry known as confinement/deconfinement phase transition. We have shown that the critical value of this phase transition, the length of strip entangling region, is decreased by increasing the momentum dissipation parameter. Again this was expected because of the decrease of the correlation length in the non-conformal vacuum.

\item Considering the holographic Wilson loop, we have shown that in theories with momentum dissipation, the correction to the potential between quark and anti quark is linear and attractive and the corresponding force between them is an increasing function of the momentum relaxation parameter.
\end{itemize}

\subsection*{Square Root Model}
Another similar family of models which leads to momentum relaxation was introduced in \cite{Taylor:2014tka}. This model is sometimes called square root model defined by the following action 
\bea\label{squareaction}
I=\int d^{d+1}x \sqrt{-g}\left[\mathcal{R}-2\Lambda-\sum_{I=1}^{d-1}\sqrt{(\partial\chi_I)^2}\right],
\eea
where $\Lambda=-\frac{d(d-1)}{2L^2}$, and $\chi_I$'s are again massless scalar fields and index $I$ runs over the spatial directions of the dual theory. This action admits the following solution for $d>2$
\begin{align}\label{ddimsquare}
\begin{split}
ds^2&=\frac{L^2}{\rho^2}\left[-f(\rho)dt^2+\frac{d\rho^2}{f(\rho)}+dx^2_{d-1}\right],\\
f(\rho)&=1-\frac{\beta}{d-1}\rho-m_0 \rho^d,\\
\chi_I(x)&=\beta \delta_{Ia}x^a.
\end{split}
\end{align}
The entropy and temperature of this solution is given by
\be
S=\frac{V_{d-1}}{4G_N}\left(\frac{\beta}{d-1}\right)^{d-1},\;\;\;\;\;\;\;\;T=\frac{\beta}{4\pi(d-1)}.
\ee
In contrast with the polynomial model, the square root model has a non-logarithmic solution for $d=2$ with
\begin{align}\label{2dimsquaref}
\begin{split}
f(\rho)&=1-\beta \rho-m_0 \rho^2,\\
\chi(x)&=\beta x,
\end{split}
\end{align}
where the above expressions are again valid for the entropy and temperature.

Here we report the result of more or less the same analysis we did for polynomial models in section \ref{sec:HEE} and \ref{sec:other} for the square root models. We will study momentum relaxation corrections to strip, spherical and cylindrical entangling regions, and we will highlight the main differences between this model and the polynomial model. These differences are all originated in the emblackening factors of these geometries.\footnote{It would be interesting to further investigate these differences from field theoretic point of view.}

For the case of strip entangling region, using the expression found in \eqref{lA1} with $f$ given by \eqref{ddimsquare}, one can easily find the correction to HEE. For $d>3$ and to the first order in $\beta$ expansion, one has
\begin{align}
\begin{split}
\Delta S=\frac{L^{d-1}H^{d-2}}{4G_N}\frac{1}{(d-1)(d-3)}& \Bigg[\frac{1}{\epsilon ^{d-3}}-\\&
\frac{ 2\sqrt{\pi} }{(d+1)\ell^{d-3}}\frac{\Gamma \left(\frac{3d-1}{2(d-1)}\right) }{\Gamma \left(\frac{d}{d-1}\right)}\left(\frac{2\sqrt{\pi}\Gamma \left(\frac{d}{2 (d-1)}\right)}{\Gamma \left(\frac{1}{2 (d-1)}\right)}\right)^{d-3}\Bigg]\beta+\mathcal{O}\left(\beta^2\right).
\end{split}
\end{align}
For the case of $d=3$, one finds
\be
\Delta S=\frac{L^{2}H}{8G_N}\left[\log\frac{\ell}{\epsilon}+\log\left(\frac{\Gamma \left(\frac{1}{4}\right)}{\sqrt{2 \pi} \Gamma \left(\frac{3}{4}\right)}\right)-\frac{1}{2}\right]\beta+\mathcal{O}\left(\beta^2\right).
\ee
The important observation is that unlike the polynomial model, here for generic $d$ (even and odd both) there is a logarithmic correction, $\log\frac{\ell}{\epsilon}$ at $\mathcal{O}\left(\beta^{d-2}\right)$. Also for $d=2$, we must use \eqref{2dimsquaref} and the corresponding HEE for $m_0=0$ is given by
\bea
\Delta S=\frac{\pi L}{4G_N}\frac{\beta\ell}{8}+\mathcal{O}\left(\beta^2\right).
\eea
Here we have found a thermal correction to the entanglement entropy in contrast to the non-critical correction we found in \eqref{eq:PolyNC} for the polynomial model. For the case of $T=0$ mixed state (extremal geometry) similar results can be found for the non-extremal case.

The area functional for the spherical and cylindrical entangling regions are given by \eqref{sphereHEE} and \eqref{cylindrHEE} respectively, where one should use $f$ from \eqref{ddimsquare} (again for the case of $m_0=0$). Here there is a crucial difference between this model and the polynomial model: the structure of subleading logarithmic correction due to momentum relaxation is shifted from even (field theory) dimensions to odd dimensions for spherical and cylindrical entangling regions. This feature is a result of the difference between the structure of the emblackening function in the solution which has a linear (in $\rho$) term instead of quadratic term which we had in the polynomial model.
For spherical entangling region at leading order in $\beta$ one has
\begin{align}
\begin{split}
\Delta S=
\begin{cases}
-\frac{\pi L^2}{8G_N}\left(\log \frac{\ell}{\epsilon}-\frac{1}{2}\right)\beta \ell, & ~~ d=3,  \\
-\frac{\pi L^3}{8G_N}\left(\frac{\ell}{\epsilon}-\frac{3\pi}{4}\right)\beta \ell, & ~~ d=4, 
\end{cases}
\end{split}
\end{align}
and for cylindrical entangling region at leading order in $\beta$ one finds
\begin{align}
\begin{split}
\Delta S=
\begin{cases}
\frac{L^2 H}{8G_N} \log \frac{\ell}{\epsilon}\left(\beta\ell\right), & ~~ d=3,  \\
\frac{\pi L^3 H}{12G_N}\left(\frac{ \ell}{\epsilon}-1\right) \left(\beta\ell\right),& ~~ d=4. 
\end{cases}
\end{split}
\end{align}
\begin{figure}\label{fig:mutual2}
\begin{center}
\includegraphics[scale=0.45]{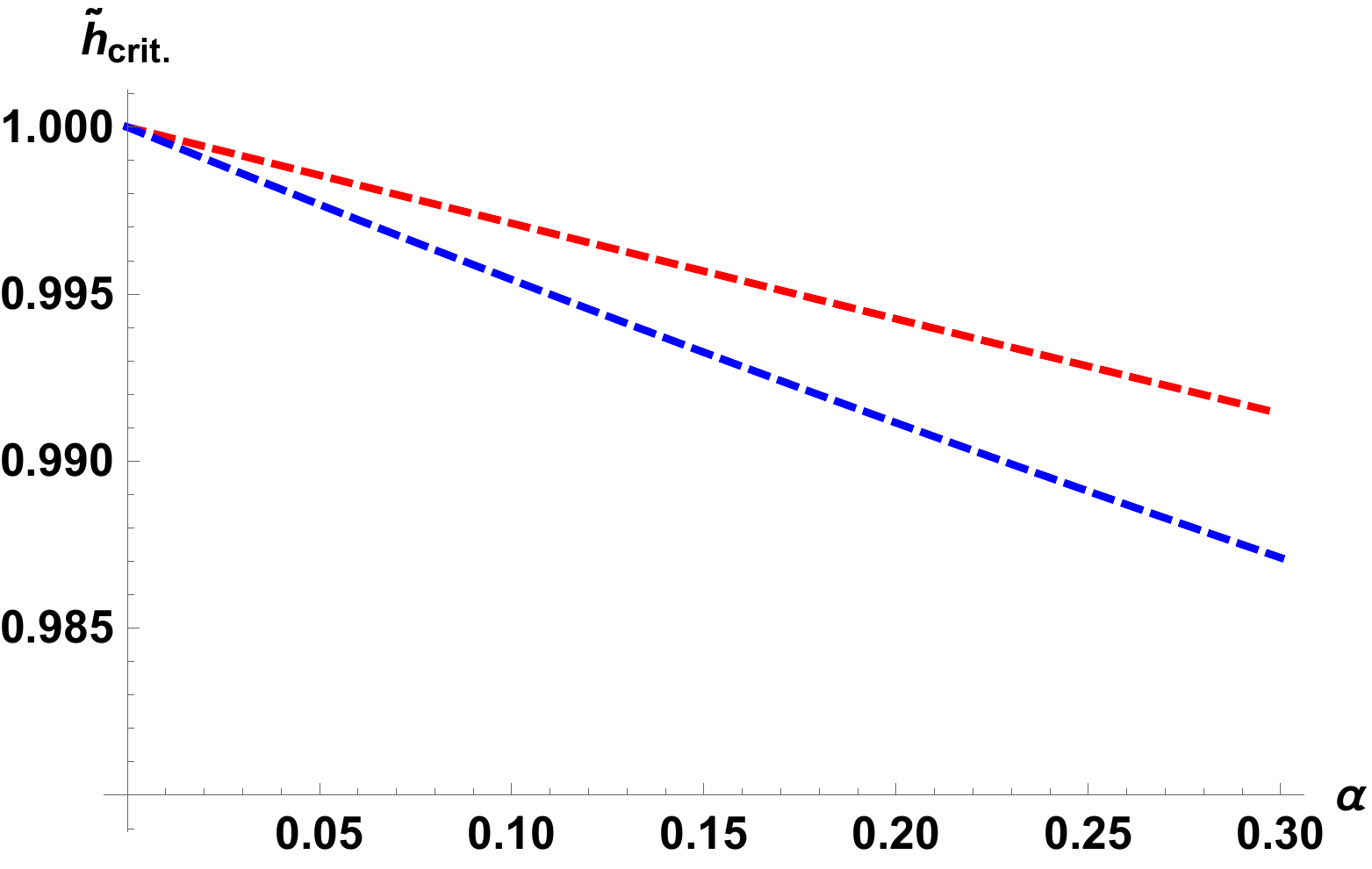}
\end{center}
\caption{Location of the critical points as a function of $\alpha$ for $d=3$ (red) and $d=4$ (blue) with $\ell=1$ in square root model. In each case, the holographic mutual information is finite below the transition curve and vanishes when we cross it.}
\end{figure}

Also in figure \ref{fig:mutual2} we have shown similar results for the phase transition of mutual information in the square root model.

\subsection*{First Law of Entanglement}
It is well-known in the literature that using the positivity of relative entropy directly implies that variation of entanglement entropy is bounded by the variation of the expectation value of modular Hamiltonian \cite{Blanco:2013joa}. In the context of holographic CFTs, this bound is saturated at first order of perturbation which is known as the first law of entanglement. First law of entanglement equates the variation of entanglement entropy for two nearby states (one is the vacuum state) with the expectation value of the modular Hamiltonian.

One may naturally ask what kind of deformed states are ought to satisfy such a law? In the context of holographic CFTs, if the reference state is the CFT ground state (the conformal vacuum in our language) and the second state is a slight deformation of it, still preserving conformal symmetry, it would be natural to expect the first law of entanglement to be satisfied. To our knowledge, this has been checked for several deformations such as relevant mass deformations in \cite{Blanco:2013joa} or for holographic duals of coherent states which are constructed with relevant or even marginal scalar deformations in \cite{Nogueira:2013if, Gentle:2013fma}.

In the case of the non-conformal vacuum one can compute for instance massive corrections to the entanglement entropy which are
\be
\Delta S=\frac{m_0\ell^2 L^{d-1}H^{d-2}}{32G_N\sqrt{\pi}}\frac{\Gamma^2(\frac{1}{2(d-1)})}{\Gamma^2(\frac{d}{2(d-1)})}\left(\frac{\Gamma(\frac{d}{d-1})}{\Gamma(\frac{d+1}{2(d-1)})}+\alpha^2\ell^2\delta_d\right)
\ee
where $\delta_d$ is a numerical factor, this is not going to be equal to the expectation value of the modular Hamiltonian (which itself is not known in this case, even for spherical or cap entangling regions).

Before finishing we would like to address possible relation between the effect of momentum relaxation in the entanglement entropy of this type of holographic states and what is well-known in the context of condensed matter physics: the central charge of critical systems changes to an effective smaller central charge due to turning on a random coupling in the system \cite{RM}.\footnote{We thank Abolhassan Vaezi for pointing out this reference to us.} It would be interesting to explore this possible relation more precisely in future works.

\section*{Acknowledgements}

We would like to thank T. Andrade, M. Faghfoor-Maghrebi,	J. Gauntlett, B. Gauxeraux, J. McGreevy, D. Tong and B. Withers for correspondence. We are grateful to S. Hartnoll, M. Rangamani, T. Takayanagi and A. Vaezi for their useful comments on an early draft. We are also grateful to M. Alishahiha for many discussions, useful comments and his collaboration during parts of this work. We thank A. Akhavan, A. Davody, A. Faraji, A. Naseh, F. Taghavi and M.R. Tanhayi for related discussions. AM would like to thank CERN-TH Division for their warm hospitality during last stages of this work. FO would like to thank school of physics of IPM for their support and warm hospitality during this project. The authors are partly supported by Iran National Science Foundation (INSF). The work of FO is also supported by the PhD student research assistant grant of Iran's National Elites Foundation.



\begin{thebibliography}{}

\bibitem{Hartnoll:2011fn} 
  S.~A.~Hartnoll,
``Horizons, holography and condensed matter,''
  arXiv:1106.4324 [hep-th].

\bibitem{Iqbal:2011ae} 
  N.~Iqbal, H.~Liu and M.~Mezei,
``Lectures on holographic non-Fermi liquids and quantum phase transitions,''
  arXiv:1110.3814 [hep-th].

\bibitem{Probe} 
  A.~Karch and A.~O'Bannon,
``Metallic AdS/CFT,''
  JHEP {\bf 0709}, 024 (2007)
  doi:10.1088/1126-6708/2007/09/024
  [arXiv:0705.3870 [hep-th]].

 S.~A.~Hartnoll, J.~Polchinski, E.~Silverstein and D.~Tong,
``Towards strange metallic holography,''
  JHEP {\bf 1004}, 120 (2010)
  doi:10.1007/JHEP04(2010)120
  [arXiv:0912.1061 [hep-th]].

 C.~Charmousis, B.~Gouteraux, B.~S.~Kim, E.~Kiritsis and R.~Meyer,
  ``Effective Holographic Theories for low-temperature condensed matter systems,''
  JHEP {\bf 1011}, 151 (2010)
  doi:10.1007/JHEP11(2010)151
  [arXiv:1005.4690 [hep-th]].
  
 B.~Gouteraux and E.~Kiritsis,
``Generalized Holographic Quantum Criticality at Finite Density,''
  JHEP {\bf 1112}, 036 (2011)
  doi:10.1007/JHEP12(2011)036
  [arXiv:1107.2116 [hep-th]].

 T.~Faulkner, N.~Iqbal, H.~Liu, J.~McGreevy and D.~Vegh,
``Strange metal transport realized by gauge/gravity duality,''
  Science {\bf 329}, 1043 (2010).
  doi:10.1126/science.1189134.  
``From Black Holes to Strange Metals,''
  arXiv:1003.1728 [hep-th].
``Charge transport by holographic Fermi surfaces,''
  Phys.\ Rev.\ D {\bf 88}, 045016 (2013)
  doi:10.1103/PhysRevD.88.045016
  [arXiv:1306.6396 [hep-th]].

\bibitem{bts1} 
  S.~A.~Hartnoll, P.~K.~Kovtun, M.~Muller and S.~Sachdev,
``Theory of the Nernst effect near quantum phase transitions in condensed matter, and in dyonic black holes,''
  Phys.\ Rev.\ B {\bf 76}, 144502 (2007)
  doi:10.1103/PhysRevB.76.144502
  [arXiv:0706.3215 [cond-mat.str-el]].

S.~A.~Hartnoll and C.~P.~Herzog,
``Impure AdS/CFT correspondence,''
  Phys.\ Rev.\ D {\bf 77}, 106009 (2008)
  doi:10.1103/PhysRevD.77.106009
  [arXiv:0801.1693 [hep-th]].

A.~Lucas, S.~Sachdev and K.~Schalm,
``Scale-invariant hyperscaling-violating holographic theories and the resistivity of strange metals with random-field disorder,''
  Phys.\ Rev.\ D {\bf 89}, no. 6, 066018 (2014)
  doi:10.1103/PhysRevD.89.066018
  [arXiv:1401.7993 [hep-th]].

\bibitem{bts2} 
 D.~Vegh,
``Holography without translational symmetry,''
  arXiv:1301.0537 [hep-th].

R.~A.~Davison,
``Momentum relaxation in holographic massive gravity,''
  Phys.\ Rev.\ D {\bf 88}, 086003 (2013)
  doi:10.1103/PhysRevD.88.086003
  [arXiv:1306.5792 [hep-th]].

M.~Blake and D.~Tong,
``Universal Resistivity from Holographic Massive Gravity,''
  Phys.\ Rev.\ D {\bf 88}, no. 10, 106004 (2013)
  doi:10.1103/PhysRevD.88.106004
  [arXiv:1308.4970 [hep-th]].

\bibitem{bts3} 
S.~A.~Hartnoll and D.~M.~Hofman,
``Locally Critical Resistivities from Umklapp Scattering,''
  Phys.\ Rev.\ Lett.\  {\bf 108}, 241601 (2012)
  doi:10.1103/PhysRevLett.108.241601
  [arXiv:1201.3917 [hep-th]].

 G.~T.~Horowitz, J.~E.~Santos and D.~Tong,
``Optical Conductivity with Holographic Lattices,''
  JHEP {\bf 1207}, 168 (2012)
  doi:10.1007/JHEP07(2012)168
  [arXiv:1204.0519 [hep-th]].

M.~Blake, D.~Tong and D.~Vegh,
``Holographic Lattices Give the Graviton an Effective Mass,''
  Phys.\ Rev.\ Lett.\  {\bf 112}, no. 7, 071602 (2014)
  doi:10.1103/PhysRevLett.112.071602
  [arXiv:1310.3832 [hep-th]].

  M.~Baggioli and O.~Pujolas,
``Electron-Phonon Interactions, Metal-Insulator Transitions, and Holographic Massive Gravity,''
  Phys.\ Rev.\ Lett.\  {\bf 114}, no. 25, 251602 (2015)
  doi:10.1103/PhysRevLett.114.251602
  [arXiv:1411.1003 [hep-th]].

  L.~Alberte, M.~Baggioli, A.~Khmelnitsky and O.~Pujolas,
``Solid Holography and Massive Gravity,''
  JHEP {\bf 1602}, 114 (2016)
  doi:10.1007/JHEP02(2016)114
  [arXiv:1510.09089 [hep-th]].

\bibitem{Andrade:2013gsa} 
  T.~Andrade and B.~Withers,
  ``A simple holographic model of momentum relaxation,''
  JHEP {\bf 1405}, 101 (2014)
  doi:10.1007/JHEP05(2014)101
  [arXiv:1311.5157 [hep-th]].

\bibitem{bts4} 
 N.~Bao, S.~Harrison, S.~Kachru and S.~Sachdev,
``Vortex Lattices and Crystalline Geometries,''
  Phys.\ Rev.\ D {\bf 88}, no. 2, 026002 (2013)
  doi:10.1103/PhysRevD.88.026002
  [arXiv:1303.4390 [hep-th]].

 N.~Bao and S.~Harrison,
``Crystalline Scaling Geometries from Vortex Lattices,''
  Phys.\ Rev.\ D {\bf 88}, 046009 (2013)
  doi:10.1103/PhysRevD.88.046009
  [arXiv:1306.1532 [hep-th]].

  M.~R.~Mohammadi Mozaffar and A.~Mollabashi,
``Crystalline geometries from a fermionic vortex lattice,''
  Phys.\ Rev.\ D {\bf 89}, no. 4, 046007 (2014)
  doi:10.1103/PhysRevD.89.046007
  [arXiv:1307.7397 [hep-th]].

L.~K.~Chen, H.~Guo and F.~W.~Shu,
``Crystalline geometries from fermionic vortex lattice with hyperscaling violation,''
  arXiv:1511.01370 [hep-th].

\bibitem{Taylor:2014tka}
  M.~Taylor and W.~Woodhead,
``Inhomogeneity simplified,''
  Eur.\ Phys.\ J.\ C {\bf 74} (2014) no.12,  3176
  doi:10.1140/epjc/s10052-014-3176-9
  [arXiv:1406.4870 [hep-th]].

\bibitem{Ge:2016lyn} 
  X.~H.~Ge, Y.~Tian, S.~Y.~Wu and S.~F.~Wu,
``Linear and quadratic in temperature resistivity from holography,''
  arXiv:1606.07905 [hep-th].

\bibitem{Roychowdhury:2015fxf} 
  D.~Roychowdhury,
  ``Holography for anisotropic branes with hyperscaling violation,''
  JHEP {\bf 1601}, 105 (2016)
  doi:10.1007/JHEP01(2016)105
  [arXiv:1511.06842 [hep-th]].
  
\bibitem{Cremonini:2016avj} 
  S.~Cremonini, H.~S.~Liu, H.~Lu and C.~N.~Pope,
  ``DC Conductivities from Non-Relativistic Scaling Geometries with Momentum Dissipation,''
  arXiv:1608.04394 [hep-th].

\bibitem{Ryu:2006bv} 
  S.~Ryu and T.~Takayanagi,
``Holographic derivation of entanglement entropy from AdS/CFT,''
  Phys.\ Rev.\ Lett.\  {\bf 96}, 181602 (2006)
  doi:10.1103/PhysRevLett.96.181602
  [hep-th/0603001].

\bibitem{Ryu:2006ef} 
  S.~Ryu and T.~Takayanagi,
``Aspects of Holographic Entanglement Entropy,''
  JHEP {\bf 0608}, 045 (2006)
  doi:10.1088/1126-6708/2006/08/045
  [hep-th/0605073].

\bibitem{Azeyanagi:2009pr} 
  T.~Azeyanagi, W.~Li and T.~Takayanagi,
``On String Theory Duals of Lifshitz-like Fixed Points,''
  JHEP {\bf 0906}, 084 (2009)
  doi:10.1088/1126-6708/2009/06/084
  [arXiv:0905.0688 [hep-th]].

\bibitem{Banerjee:2014oaa} 
  S.~Banerjee, A.~Bhattacharyya, A.~Kaviraj, K.~Sen and A.~Sinha,
``Constraining gravity using entanglement in AdS/CFT,''
  JHEP {\bf 1405}, 029 (2014)
  doi:10.1007/JHEP05(2014)029
  [arXiv:1401.5089 [hep-th]].

\bibitem{Calabrese:2004eu} 
  P.~Calabrese and J.~L.~Cardy,
``Entanglement entropy and quantum field theory,''
  J.\ Stat.\ Mech.\  {\bf 0406}, P06002 (2004)
  doi:10.1088/1742-5468/2004/06/P06002
  [hep-th/0405152].

\bibitem{LARGE} 
W.~Fischler and S.~Kundu,
``Strongly Coupled Gauge Theories: High and Low Temperature Behavior of Non-local Observables,''
  JHEP {\bf 1305}, 098 (2013)
  doi:10.1007/JHEP05(2013)098
  [arXiv:1212.2643 [hep-th]].

W.~Fischler, A.~Kundu and S.~Kundu,
``Holographic Mutual Information at Finite Temperature,''
  Phys.\ Rev.\ D {\bf 87}, no. 12, 126012 (2013)
  doi:10.1103/PhysRevD.87.126012
  [arXiv:1212.4764 [hep-th]].

S.~Kundu and J.~F.~Pedraza,
``Aspects of Holographic Entanglement at Finite Temperature and Chemical Potential,''
  arXiv:1602.07353 [hep-th].
  
\bibitem{Solodukhin:2008dh} 
  S.~N.~Solodukhin,
  ``Entanglement entropy, conformal invariance and extrinsic geometry,''
  Phys.\ Lett.\ B {\bf 665}, 305 (2008)
  doi:10.1016/j.physletb.2008.05.071
  [arXiv:0802.3117 [hep-th]].

\bibitem{Hung:2011} 
L.-Y. Hung, R. C. Myers, and M. Smolkin, On Holographic Entanglement Entropy and Higher
Curvature Gravity," JHEP 04 (2011) 025, 1101.5813.

\bibitem{Singh:2012ala} 
  A.~Singh,
``Holographic Entanglement Entropy: RG Flows and Singular Surfaces,''
PhD thesis, University of Waterloo, 2012.

\bibitem{Myers:2012ed} 
  R.~C.~Myers and A.~Singh,
  ``Comments on Holographic Entanglement Entropy and RG Flows,''
  JHEP {\bf 1204}, 122 (2012)
  doi:10.1007/JHEP04(2012)122
  [arXiv:1202.2068 [hep-th]].

\bibitem{FS} 
  B.~Swingle,
``Entanglement Entropy and the Fermi Surface,''
  Phys.\ Rev.\ Lett.\  {\bf 105}, 050502 (2010)
  doi:10.1103/PhysRevLett.105.050502
  [arXiv:0908.1724 [cond-mat.str-el]].

 Y.~Zhang, T.~Grover and A.~Vishwanath,
``Entanglement entropy of critical spin liquids,''
  Phys.\ Rev.\ Lett.\  {\bf 107}, 067202 (2011)
  doi:10.1103/PhysRevLett.107.067202
  [arXiv:1102.0350 [cond-mat.str-el]].

W. Ding, A. Seidel, and K. Yang,
"Entanglement Entropy of Fermi Liquids via Multidimensional Bosonization,"
  [arXiv:1110.3004 [cond-mat.stat-mech]].

 M.~M.~Wolf,
``Violation of the entropic area law for Fermions,''
  Phys.\ Rev.\ Lett.\  {\bf 96}, 010404 (2006)
  doi:10.1103/PhysRevLett.96.010404
  [quant-ph/0503219].

T. Barthel, M.-C. Chung, and U. Schollwock,
""Entanglement scaling in critical two dimensional fermionic and bosonic systems,"
Phys.\ Rev.\ A\ {\bf 74}, 022329 (2006)
[arXiv:condmat/0602077].

W. Li, L. Ding, R. Yu, T. Roscilde, and S. Haas,
"Scaling Behavior of Entanglement in Two- and Three-Dimensional Free Fermions,"
Phys.\ Rev.\ B\ {\bf 74}, 073103 (2006)
[arXiv:quantph/0602094].

P.~Calabrese, M.~Mintchev and E.~Vicari,
``Entanglement entropies in free fermion gases for arbitrary dimension,''
  Europhys.\ Lett.\  {\bf 97}, 20009 (2012)
  doi:10.1209/0295-5075/97/20009
  [arXiv:1110.6276 [cond-mat.quant-gas]].

\bibitem{Ogawa:2011bz} 
  N.~Ogawa, T.~Takayanagi and T.~Ugajin,
``Holographic Fermi Surfaces and Entanglement Entropy,''
  JHEP {\bf 1201}, 125 (2012)
  doi:10.1007/JHEP01(2012)125
  [arXiv:1111.1023 [hep-th]].

\bibitem{Huijse:2011ef} 
  L.~Huijse, S.~Sachdev and B.~Swingle,
``Hidden Fermi surfaces in compressible states of gauge-gravity duality,''
  Phys.\ Rev.\ B {\bf 85}, 035121 (2012)
  doi:10.1103/PhysRevB.85.035121
  [arXiv:1112.0573 [cond-mat.str-el]].

\bibitem{Takayanagi:2013ywa} 
  T.~Takayanagi,
``Strange metals and holographic entanglement entropy,''
  Int.\ J.\ Mod.\ Phys.\ A {\bf 28}, 1340004 (2013).
  doi:10.1142/S0217751X13400046

\bibitem{Alishahfonda} 
  M.~Alishahiha, A.~F.~Astaneh and M.~R.~Mohammadi Mozaffar,
``Thermalization in backgrounds with hyperscaling violating factor,''
  Phys.\ Rev.\ D {\bf 90}, no. 4, 046004 (2014)
  doi:10.1103/PhysRevD.90.046004
  [arXiv:1401.2807 [hep-th]], 
  P.~Fonda, L.~Franti, V.~Keränen, E.~Keski-Vakkuri, L.~Thorlacius and E.~Tonni,
``Holographic thermalization with Lifshitz scaling and hyperscaling violation,''
  JHEP {\bf 1408}, 051 (2014)
  doi:10.1007/JHEP08(2014)051
  [arXiv:1401.6088 [hep-th]].

\bibitem{Hosseini:2015gua} 
  S.~M.~Hosseini and Á.~Véliz-Osorio,
``Entanglement and mutual information in two-dimensional nonrelativistic field theories,''
  Phys.\ Rev.\ D {\bf 93}, no. 2, 026010 (2016)
  [Phys.\ Rev.\ D {\bf 93}, 026010 (2016)]
  doi:10.1103/PhysRevD.93.026010
  [arXiv:1510.03876 [hep-th]].

\bibitem{Basanisi:2016hsh} 
  L.~Basanisi and S.~Chakrabortty,
``Holographic Entanglement Entropy in NMG,''
  arXiv:1606.01920 [hep-th].

\bibitem{Ling:2016ien} 
  Y.~Ling, Z.~Y.~Xian and Z.~Zhou,
``Holographic Shear Viscosity in Hyperscaling Violating Theories without Translational Invariance,''
  arXiv:1605.03879 [hep-th].

\bibitem{Headrick:2010zt} 
  M.~Headrick,
``Entanglement Renyi entropies in holographic theories,''
  Phys.\ Rev.\ D {\bf 82}, 126010 (2010)
  doi:10.1103/PhysRevD.82.126010
  [arXiv:1006.0047 [hep-th]].

\bibitem{Mozaffar:2015xue} 
  M.~R.~Mohammadi Mozaffar, A.~Mollabashi and F.~Omidi,
``Holographic Mutual Information for Singular Surfaces,''
  JHEP {\bf 1512}, 082 (2015)
  doi:10.1007/JHEP12(2015)082
  [arXiv:1511.00244 [hep-th]].

\bibitem{Hayden:2011ag} 
  P.~Hayden, M.~Headrick and A.~Maloney,
``Holographic Mutual Information is Monogamous,''
  Phys.\ Rev.\ D {\bf 87}, no. 4, 046003 (2013)
  doi:10.1103/PhysRevD.87.046003
  [arXiv:1107.2940 [hep-th]].

\bibitem{npar} 
  M.~Alishahiha, M.~R.~Mohammadi Mozaffar and M.~R.~Tanhayi,
``On the Time Evolution of Holographic n-partite Information,''
  JHEP {\bf 1509}, 165 (2015)
  doi:10.1007/JHEP09(2015)165
  [arXiv:1406.7677 [hep-th]].
  
  M.~R.~Tanhayi,
``Thermalization of Mutual Information in Hyperscaling Violating Backgrounds,''
  JHEP {\bf 1603}, 202 (2016)
  doi:10.1007/JHEP03(2016)202
  [arXiv:1512.04104 [hep-th]].

  S.~Mirabi, M.~R.~Tanhayi and R.~Vazirian,
``On the Monogamy of Holographic $n$-partite Information,''
  Phys.\ Rev.\ D {\bf 93}, no. 10, 104049 (2016)
  doi:10.1103/PhysRevD.93.104049
  [arXiv:1603.00184 [hep-th]].
  
\bibitem{fidelity0}
S.~L.~Braunstein and C.~M.~Caves,
``Statisical distance and the geometry of quantum states",
Phys.\ Rev.\ Lett.\ {\bf 72} (1994) 3439.

\bibitem{fidelity}
S-J.~Gu
``Fidelity approach to quantum phase transitions," 
Int.\ J.\ Mod.\ Phys.\ B 24, 4371 (2010)
[arXiv:0811.3127 [quant-ph]].

\bibitem{MIyaji:2015mia} 
  M.~Miyaji, T.~Numasawa, N.~Shiba, T.~Takayanagi and K.~Watanabe,
``Distance between Quantum States and Gauge-Gravity Duality,''
  Phys.\ Rev.\ Lett.\  {\bf 115}, no. 26, 261602 (2015)
  doi:10.1103/PhysRevLett.115.261602
  [arXiv:1507.07555 [hep-th]].
    
\bibitem{Bak:2015jxd} 
  D.~Bak,
 ``Information metric and Euclidean Janus correspondence,''
  Phys.\ Lett.\ B {\bf 756}, 200 (2016)
  doi:10.1016/j.physletb.2016.03.012
  [arXiv:1512.04735 [hep-th]].

\bibitem{Alishahiha:2015rta} 
  M.~Alishahiha,
``Holographic Complexity,''
  Phys.\ Rev.\ D {\bf 92}, no. 12, 126009 (2015)
  doi:10.1103/PhysRevD.92.126009
  [arXiv:1509.06614 [hep-th]].

\bibitem{Nishioka:2006gr} 
T.~Nishioka and T.~Takayanagi,
``AdS Bubbles, Entropy and Closed String Tachyons,''
JHEP {\bf 0701}, 090 (2007)
doi:10.1088/1126-6708/2007/01/090
[hep-th/0611035].

\bibitem{Klebanov:2007ws} 
  I.~R.~Klebanov, D.~Kutasov and A.~Murugan,
``Entanglement as a probe of confinement,''
  Nucl.\ Phys.\ B {\bf 796}, 274 (2008)
  doi:10.1016/j.nuclphysb.2007.12.017
  [arXiv:0709.2140 [hep-th]].

\bibitem{Nishioka:2009un} 
T.~Nishioka, S.~Ryu and T.~Takayanagi,
``Holographic Entanglement Entropy: An Overview,''
J.\ Phys.\ A {\bf 42}, 504008 (2009)
doi:10.1088/1751-8113/42/50/504008
[arXiv:0905.0932 [hep-th]].

\bibitem{Maldacena:1998im} 
  J.~M.~Maldacena,
``Wilson loops in large N field theories,''
  Phys.\ Rev.\ Lett.\  {\bf 80}, 4859 (1998)
  doi:10.1103/PhysRevLett.80.4859
  [hep-th/9803002].

\bibitem{Giataganas:2012zy} 
  D.~Giataganas,
  ``Probing strongly coupled anisotropic plasma,''
  JHEP {\bf 1207}, 031 (2012)
  doi:10.1007/JHEP07(2012)031
  [arXiv:1202.4436 [hep-th]].
  
\bibitem{Blanco:2013joa} 
  D.~D.~Blanco, H.~Casini, L.~Y.~Hung and R.~C.~Myers,
``Relative Entropy and Holography,''
  JHEP {\bf 1308}, 060 (2013)
  doi:10.1007/JHEP08(2013)060
  [arXiv:1305.3182 [hep-th]].

J.~Bhattacharya, M.~Nozaki, T.~Takayanagi and T.~Ugajin,
  ``Thermodynamical Property of Entanglement Entropy for Excited States,''
  Phys.\ Rev.\ Lett.\  {\bf 110}, no. 9, 091602 (2013)
  doi:10.1103/PhysRevLett.110.091602
  [arXiv:1212.1164].
  
G.~Wong, I.~Klich, L.~A.~Pando Zayas and D.~Vaman,
``Entanglement Temperature and Entanglement Entropy of Excited States,''
  JHEP {\bf 1312}, 020 (2013)
  doi:10.1007/JHEP12(2013)020
  [arXiv:1305.3291 [hep-th]].  
  
D.~Allahbakhshi, M.~Alishahiha and A.~Naseh,
``Entanglement Thermodynamics,''
  JHEP {\bf 1308}, 102 (2013)
  doi:10.1007/JHEP08(2013)102
  [arXiv:1305.2728 [hep-th]].

 N.~Lashkari, M.~B.~McDermott and M.~Van Raamsdonk,
``Gravitational dynamics from entanglement 'thermodynamics',''
  JHEP {\bf 1404}, 195 (2014)
  doi:10.1007/JHEP04(2014)195
  [arXiv:1308.3716 [hep-th]].  

\bibitem{Nogueira:2013if} 
  F.~Nogueira,
``Extremal Surfaces in Asymptotically AdS Charged Boson Stars Backgrounds,''
  Phys.\ Rev.\ D {\bf 87}, 106006 (2013)
  doi:10.1103/PhysRevD.87.106006
  [arXiv:1301.4316 [hep-th]].

\bibitem{Gentle:2013fma} 
  S.~A.~Gentle and M.~Rangamani,
``Holographic entanglement and causal information in coherent states,''
  JHEP {\bf 1401}, 120 (2014)
  doi:10.1007/JHEP01(2014)120
  [arXiv:1311.0015 [hep-th]].

\bibitem{RM} 
G.~Refael and J.~E.~Moore
``Entanglement entropy of random quantum critical points in one dimension,''
  Phys.\ Rev.\ Lett.\  {\bf 93}, 260602 (2004)
  [arXiv:0406737 [con-mat]].

\end{thebibliography}
\end{document}